\documentclass[]{IEEEtran}
\IEEEoverridecommandlockouts

\ifCLASSINFOpdf
\else
\fi

\usepackage{multirow}
\usepackage{cite}
\usepackage{stfloats}
\usepackage{color}
\usepackage{graphicx}
\usepackage{epstopdf}
\usepackage{slashbox}
\usepackage{float}
\usepackage{amssymb }
\usepackage[cmex10]{amsmath}
\usepackage{algorithm} 
\usepackage{algorithmic}
\usepackage{setspace}
\usepackage{amsfonts}
\usepackage{amsmath}
\usepackage{booktabs}
\usepackage{amsmath,bm}
\usepackage{tabularx}
\usepackage{makecell}
\usepackage{subfigure}

\def\BibTeX{{\rm B\kern-.05em{\sc i\kern-.025em b}\kern-.08em
    T\kern-.1667em\lower.7ex\hbox{E}\kern-.125emX}}
\usepackage{balance}
\begin{document}
\title{Resource Allocation for Mutualistic Symbiotic Radio with Hybrid Active-Passive Communications}
\author{\IEEEauthorblockN{Hong Guo, Yinghui Ye,~\IEEEmembership{Member,~IEEE}, Haijian Sun,~\IEEEmembership{Member,~IEEE}, \\Liqin Shi, and Rose Qingyang Hu,~\IEEEmembership{Fellow,~IEEE}}
\thanks{ Hong Guo, Yinghui Ye, and Liqin Shi are with the Shaanxi Key Laboratory of Information Communication Network and Security, Xi'an University of Posts \& Telecommunications, China (e-mail: hongguo@stu.xupt.edu.cn, connectyyh@126.com, liqinshi@hotmail.com).}
\thanks{Haijian Sun is with the School of Electrical and Computer Engineering, The University of Georgia, Athens, GA, USA (email: hsun@uga.edu).}
\thanks{ Rose Qingyang Hu is with the Department of Electrical and Computer Engineering at Utah State University, USA (e-mail: rose.hu@usu.edu).}
\thanks{This work was presented in part at IEEE Global Communications Conference (Globecom), Kuala Lumpur, Malaysia, Dec., 2023 [6].}
}
\maketitle

\begin{abstract}
Mutualistic symbiotic radio (SR) is a communication paradigm that offers high spectrum efficiency and low power consumption, where the secondary user (SU) transmits information by modulating and backscattering the primary transmitter's (PT's) signal, enabling shared use of spectrum and power with PT. In return, the PT's performance can be enhanced by SU's backscattered signal, forming a mutualistic relationship. However, the low modulation rate causes extremely inferior transmission rates for SUs. To improve the SU transmission rate, this paper proposes a new mutualistic SR with hybrid active-passive communications (HAPC) to explore the tradeoff between backscatter communications (BC) and active communications (AC) in terms of power consumption and transmission rate, enabling each SU to transmit signal via passive BC and  AC alternatively. We propose two problems to maximize the total rate of all SUs under the fixed and dynamic successive interference cancellation (SIC) ordering, respectively. The fixed SIC ordering-based problem is to jointly optimize the SUs' reflection coefficients, the transmit power of each SU during AC, and the time allocation for each SU during BC and AC, subject to the energy causality constraint and the PT's transmission rate gain constraint. In addition to pondering the constraints involved in the fixed SIC ordering-based problem, the dynamic SIC ordering-based problem, which is a mixed integer programming one,  further considers the SIC ordering constraint.
 The above two problems are solved by our proposed successive convex approximation (SCA)-based and block coordinate descent (BCD)-based iterative algorithms, respectively.
  Simulation results demonstrate that: 1) the proposed mutualistic SR system outperforms traditional designs in terms of the rates achieved by SUs under the same constraints;  2) the total rate of all SUs under the dynamic SIC ordering is larger than that of the fixed one when the PT's minimum rate gain is high, and becomes nearly identical when the PT's minimum rate gain is low.
\end{abstract}

\IEEEpeerreviewmaketitle
\section{Introduction}
\IEEEPARstart{A}s reported by \cite{8879484}, around 80 billion Internet-of-Things (IoT) devices are expected to be connected by 2030. With such a massive deployment of IoT devices, two challenges arise. On the one hand, supporting such massive connections requires a substantial amount of available radio spectrum resources. However, the majority of suitable spectrum for IoTs has already been allocated, leading to a severe scarcity of available radio resources. On the other hand, some IoT devices deployed in harsh scenarios, such as high temperatures, high pressures, or toxic environments, are required to support continuous operation up to 10-20 years \cite{9475506}, which poses an extremely high demand on battery life. Hence, there is an urgent need for high spectrum-efficient and extremely low-power IoT devices \cite{10130082}.

To address this challenge, symbiotic radio (SR\footnote{ SR outperforms traditional BC paradigms, such as monostatic, bistatic, and ambient BC, by enabling flexible deployment, avoiding dedicated RF sources, and mitigating direct-link interference. }) has been proposed by allowing IoT devices (also termed as secondary users (SUs) in this paper) to share both the spectrum and power resources with primary transmitters (PTs), e.g., cellular transmitters, via passive backscatter communications (BC) \cite{10437703,8907447,9193946,9749195,11020590}. Specifically, the SUs modulate and backscatter the primary signal to the associated receiver by switching their load impedances, rather than generate carrier by themselves, leading to an extra-low power consumption. Meanwhile, SUs harvest energy from the primary signals to compensate for their own circuit consumptions.
SR can be generally classified into three categories based on the interaction between the PT and SU: commensal, parasitic, and mutualistic SR \cite{9749195, 8692391}. In commensal SR, the SU backscatters the primary signal without affecting the PT, i.e., the interference from the SU to the PT is negligible. In parasitic SR, the SU maximizes its own data rate at the cost of degrading the PT's performance. In mutualistic SR, the SU's backscattered signal can enhance the PT's transmission performance. This performance gain arises from the significantly longer symbol duration of the SU, which enables its backscattered signal to function as a constructive multipath component. Owing to the above advantage, we focus on mutualistic SR in this paper, and relevant works are reviewed as follows.

Recent studies have explored mutualistic SR from the perspective of resource allocation. Works in \cite{8907447,9866050,8665892} studied the weighted sum rate problem. Specifically, the work in \cite{8907447} optimized PT's transmit beamforming to maximize the weighted sum rate and minimize the PT's transmit power, respectively. \cite{9866050} considered the hardware impairments in the mutualistic SR, and maximized the weighted sum rate by jointly optimizing the PT's transmit power and the SU's reflection coefficient. A similar optimization problem was studied in \cite{8665892} while considering the effect of channel fading. In \cite{8692391}, the PT's average transmission rate was maximized by considering its average power constraint. Under the PT's transmit power constraint and the achievable throughput requirements for both PT and SU, Chu \emph{et al.} proposed an optimization problem to maximize the system's energy efficiency \cite{9036977}. Different from previous works in \cite{8907447,9866050,8692391,8665892,9036977} with only one SU, \cite{9686018,9461158} considered multiple SUs in mutualistic SR. To be specific, Xu \emph{et al.} optimized the receiver's beamforming vector to maximize the PT's rate \cite{9686018}. Yang \emph{et al.} proposed an energy efficiency maximization problem by jointly optimizing the PT's transmit power, the SUs' reflection coefficients, and the time allocated to each SU for backscattering, subject to the energy-causality constraint of each SU \cite{9461158}.

In addition to the above-mentioned works on resource allocation in mutualistic SR, some studies have also investigated it from the perspective of performance analysis. The authors in \cite{9600844} derived the channel capacity of a mutualistic SR system into the closed form under the SU's circuit-sensitive constraint. Similarly, \cite{10214560} investigated the capacity performance while considering the sensitivity constraint at the SU and provided an analytical expression of the ergodic sum capacity of PT and SU. The work in \cite{10130794} extended the single-SU scenario considered in \cite{9600844,10214560} to a multi-SU scenario, and derived closed-form expressions for the ergodic rates and outage probabilities of both the PT and the SUs.
 The work in \cite{10107747} considered a more general model in which several multi-antenna PTs were deployed to serve multiple SUs and receivers, and derived expressions for the average signal-to-interference-plus-noise ratios of both the PTs and the SUs.

Although previous studies \cite{8907447,9866050,8692391,8665892,9036977,9686018,9461158,9600844,10214560,10130794,10107747} have demonstrated the superior performance of mutualistic SR, especially in terms of low-power consumption, the SU's transmission rate is still low due to the large modulation rate difference\footnote{Since the symbol duration of SU is much longer than that of PT, the modulation rate of SU is therefore much lower than that of PT, resulting in a modulation rate difference.} between PT and SU when SU transmits signals via passive BC, and this causes SU to have an equivalent spread-spectrum code of length \emph{N} (see (18) in \cite{8907447}. Such a drawback hinders the application of mutualistic SR in IoT. For example, the frequency modulation backscatter can achieve a rate up to 3.2 kbps at ranges of 1.54$-$18 m \cite{wang2017fm} and that Wi-Fi based BC achieves a data rate of up to 1 kbps with a range up to 14 m \cite{zhang2016hitchhike}. Both fall below the 3rd generation partnership project (3GPP)'s 5 kbps maximum data rate that is required for smart city/farm and so on \cite{10463656}.

In contrast, active communication (AC) achieves high data rates by generating its own RF carrier but suffers from high energy consumption due to the use of active components, such as oscillators. Thus, there is a tradeoff between AC and BC in terms of power consumption and communication rate, which motivates us to consider the mutualistic SR with hybrid active-passive communications (HAPC)\footnote{Existing works on HAPC either do not consider spectrum reuse \cite{8340034}, \cite{8957679} or fail to achieve the mutualistic transmission between PT and SU \cite{9253590}, \cite{8327597}, and are therefore fundamentally different from our framework.} to improve SUs' transmission rate\footnote{Though the active-load BC can improve SU's rate by amplifying incident signals \cite{9154299}, it cannot overcome the large modulation rate difference between PT and SU, which is a key factor limiting the further increase of SU's rate in mutualistic SR.}. In our proposed framework, SUs can dynamically perform both BC and AC using harvested energy, rather than being limited to passive BC as in existing mutualistic SR.  Moreover, our proposed mutualistic SR with HAPC is designed to ensure a mutualistic relationship between the PT and SUs, which cannot be achieved in existing HAPC networks. The main contributions of this paper are summarized as follows.

\begin{itemize}
\item We propose a novel mutualistic SR with HAPC to overcome the rate limitation of SUs in traditional mutualistic SR. By allowing each SU to perform both BC and AC, two new tradeoffs arise: 1) a tradeoff between BC and AC in terms of energy consumption and transmission rate; and 2) a tradeoff between the PT's rate gain and the SUs' rate improvement. These tradeoffs not only affect the successive interference cancellation (SIC) ordering when SUs perform AC, but also result in tightly coupled optimization variables and highly non-convex objective functions and constraints, which poses new challenges to the design joint resource allocation strategies.
\item We first consider a fixed SIC ordering scheme in the proposed mutualistic SR, and explore the optimal tradeoff between BC and AC. To this end, we formulate a problem for maximizing SUs' total rate by jointly optimizing the SUs' reflection coefficients, the transmit power of each SU during AC, and the time allocation of each SU for BC and AC, while ensuring the mutualistic relationship between PT and SUs as well as the energy causality constraints of SUs. The problem is non-convex due to the presence of multiple coupled variables in the objective function and the constraints. Using successive convex approximation (SCA) and auxiliary variables, we transform the original problem into a convex one that can be solved by the proposed SCA-based iterative algorithm.
\item We further introduce a dynamic SIC ordering scheme to enhance SUs' transmission rate, especially when the PT's minimum rate gain, i.e., the sum of the rate gain during BC and the rate loss during AC, is high. This setting allows flexible decoding order at the receiver to better balance the  tradeoff between the PT's and SUs' rates.
    Under this setting, we formulate a mixed integer programming (MIP)  problem for maximizing the SUs' total rates by jointly optimizing the SUs' reflection coefficients, the transmit power of each SU during AC, the time allocation for each SU between BC and AC, and the SIC ordering. To address the non-convexity, we firstly use a block coordinate descent (BCD) technique to decompose the original problem into two sub-problems. The first sub-problem optimizes the SUs' reflection coefficients, the transmit power of each SU during AC, the time allocation for each SU between BC and AC with the given SIC ordering, and the second one optimizes the SIC ordering with the obtained optimization variables. Both subproblems are transformed to convex ones and then a BCD-based iterative algorithm is proposed to obtain resource allocation results under the dynamic SIC ordering.

\item Simulation results demonstrate the following three aspects. First, the proposed SCA- and BCD-based  algorithms exhibit a fast convergence speed. Second, the proposed scheme achieves a significantly higher total rate for the SUs compared to traditional mutualistic SR, regardless of whether fixed or dynamic SIC ordering is adopted.  Third, compared to the fixed SIC ordering, the dynamic one achieves the same rate of all SUs when the PT's minimum rate gain is low, while obtains a higher rate of all SUs when the minimum rate gain is large.
\end{itemize}

The rest of the paper is organized as follows. The system model of mutualistic SR with HAPC is presented in Section II. In Section III, the SUs' total rate maximization problem with fixed SIC ordering is formulated and solved. Section IV formulates and solves the SUs' total rate optimization problem with dynamic SIC ordering. Section V analyzes the computational complexity and convergence of the proposed algorithms. Section VI presents the numerical results, and Section VII concludes the paper.

\section{System Model}
 The proposed mutualistic SR with HAPC consists of a single-antenna PT, $\emph{K}$ single-antenna SUs and a single-antenna receiver, as illustrated in Fig. 1. Each SU is equipped with four separate circuits, namely the BC circuit, the AC circuit, the energy harvesting (EH) circuit and the energy storage circuit.
     \begin{table}[h!]
  \begin{center}
    \caption{ Definitions of notations}
    \small
    \begin{tabular}{|l|l|}
     \hline
      \textbf{Notation} & \textbf{Definitions} \\
      \hline
      $s(n)$ & The PT's signal\\
      \hline
      $W$ &  The channel bandwidth\\
      \hline
      $\emph{T}_b$ &  The BC time of all SUs\\
      \hline
      $\emph{T}_a$ &  The AC time of all SUs\\
      \hline
      ${P_p}$ & The transmit power of PT\\
      \hline
      ${{P_{tr\emph{i}}}}$ & The transmit power of $\text{SU}_{\emph{i}}$\\
      \hline
      $c_\emph{i}$ & $\text{SU}_{\emph{i}}$'s signal during BC phase\\
      \hline
      ${x_\emph{i}(n)}$ & $\text{SU}_{\emph{i}}$'s signal during AC phase\\
      \hline
      ${\beta _\emph{i}}$ & The reflection coefficient of $\text{SU}_{\emph{i}}$\\
      \hline
      $\eta$ & The energy harvesting coefficient\\
      \hline
      $w\left( n \right)$ & The additive white Gaussian noise\\
      \hline
      $\tau_{\emph{i}}$ & The time allocated to the $\text{SU}_{\emph{i}}$ for BC\\
      \hline
      $t_{\emph{i}}$ & The time allocated to the $\text{SU}_{\emph{i}}$ for AC\\
      \hline
      ${\Delta _1}$ & SUs' increased rate during BC phase\\
      \hline
      ${\Delta _2}$ & SUs' decreased rate during AC phase\\
      \hline
       ${\varepsilon _b}$ & The static circuit power consumption of BC\\
      \hline
      $\varepsilon _a$ & The  static circuit power consumption of AC\\
      \hline
      ${g_{\rm{PS},\emph{i}}}$ &  The channel coefficient of the ${\rm{PT}} \to \text{SU}_{\emph{i}} \; \rm{link}$\\
      \hline
      ${g_{\rm{SS},\emph{ij}}}$ & The channel coefficient of the $\text{SU}_{\emph{i}} \to \text{SU}_{\emph{j}} \; \rm{link}$ \\
      \hline
      ${g_{\rm{PR}}}$ &  The channel coefficient of the ${\rm{PT}} \to $ Receiver \rm{link}\\
      \hline
      ${g_{\rm{SR,\emph{i}}}}$ & The channel coefficient of the $\text{SU}_{\emph{i}} \to $ Receiver \rm{link}\\
      \hline
    \end{tabular}
  \end{center}
\end{table}
 The duration of an entire transmission block equals ${\emph{T}}$, which is the sum of the BC time of all SUs, i.e., ${\emph{T}_b}$, and the AC time of all SUs, i.e., ${\emph{T}_a}$. Without loss of generality, we assume ${\emph{T}}=1$. All SUs operate in the time division multiple access (TDMA) mode. Both  ${\emph{T}_b}$ and ${\emph{T}_a}$ are divided into $\emph{K}$ time slots. Denote the time allocated to the $\text{SU}_{\emph{i}}$ for BC and AC as ${\tau _\emph{i}}$ and ${t_\emph{i}}$, respectively. We have $\sum\nolimits_{\emph{i} = 1}^{\emph{K}} {{\tau _\emph{i}}}  \le {\emph{T}_b}$, $\sum\nolimits_{\emph{i} = 1}^{\emph{K}} {{t_\emph{i}}}  \le {\emph{T}_a}$. To enhance energy efficiency, each SU is designed to harvest energy not only from the PT's signal but also from other SUs' transmissions during both BC and AC phases, thereby enabling energy recycling.

Denote ${g_{\rm{PR}}}$, ${g_{\rm{PS},\emph{i}}}$, ${g_{\rm{SS},\emph{ij}}}$, and ${g_{\rm{SR,\emph{i}}}}$ as the channel coefficients of the link between the PT and the receiver, the link between the PT and the $\text{SU}_{\emph{i}}$, the link between the $\text{SU}_{\emph{i}}$ and the $\text{SU}_{\emph{j}}$, and the link from the $\text{SU}_{\emph{i}}$ to the receiver, respectively. A block flat-fading channel model is considered, meaning that all the above channel coefficients remain unchanged within a transmission block but  vary  across different blocks. Moreover, we assume perfect synchronization and channel state information\footnote{Please note that the presented analysis can be extended to imperfect SIC and partial CSI.} (CSI) at the receiver, as they are assumed to be obtained prior to the transmission phase and thus their overhead is not included in the system time slot structure. The key notations are listed in Table I.

\begin{figure}
  \centering
  \includegraphics[width=0.4\textwidth]{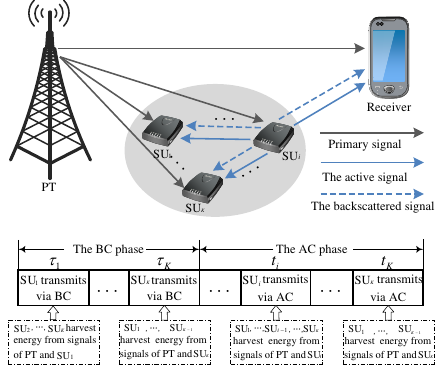}\\
  \caption{System model and frame structure.}
\end{figure}

\subsection{Passive Backscatter Communication Phase}
During  the BC period ${\emph{T}_b}$, each SU sequentially transmits signals via passive BC. For example, within $\tau_\emph{i}$, the $\text{SU}_{\emph{i}}$ backscatters its own information ${c_\emph{i}}$ to the receiver. The backscattered symbol spans \emph{N} primary symbol periods with $\emph{N} \gg 1$. Meanwhile, $\text{SU}_{\emph{i}}$ harvests energy from the primary signal, and the other SUs harvest energy from both the PT's signal and the backscattered signal. From the information transmission's perspective, the receiver receives both the backscattered signal and the primary signal simultaneously, and its received signal ${y_{b\emph{i}}}\left( n \right)$ can be written as \vspace{-4pt}
\begin{align}\label{1}
{y_{b\emph{i}}}\left( n \right) &= \underbrace {\sqrt {{P_p}} {g_{\rm{PR}}}s\left( n \right)}_{{\rm{PT}} \to {\rm{receiver\; link}}} + \underbrace {\sqrt {{P_p}} \sqrt {{\beta _\emph{i}}} {c_\emph{i}} {g_{\rm{PS},\emph{i}}}{g_{\rm{SR},\emph{i}}}s(n)}_{{\rm{a\; multipath\; component\; of\; s(n)}}} + w\left( n \right),
\end{align}
where ${P_p}$ is the transmit power of the PT, ${\beta _\emph{i}}$ denotes the $\text{SU}_{\emph{i}}$'s reflection coefficient, $s(n)$, $n = 1,...,\emph{N}$, denotes the primary signal transmitted by the PT, which follows a standard circularly symmetric complex Gaussian distribution, and $w(n)$ is the additive white Gaussian noise (AWGN) with mean zero and power ${\sigma ^2}$.

The second term in \eqref{1} can be regarded as the output
of the primary signal $s(n)$ passing through a slowly varying channel $\sqrt {{\beta _\emph{i}}} {c_\emph{i}} {g_{\rm{PS},\emph{i}}}{g_{\rm{SR},\emph{i}}}$. Thus, the receiver first decodes the primary signal $s(n)$ by treating the backscattered signal as a multipath component \cite{9751388}. The rate of PT within ${\tau _\emph{i}}$ is given as
\begin{align}\label{2}
R_p^{{\tau _\emph{i}}}\! &=\! W \! {\tau _\emph{i}}  {\mathbb{E}_{c_\emph{i}}}\!\!\left[ {{{\log }_2}\!\left( {1\! +\! \frac{{{P_p}{{\left| {{g_{{\rm{PR}}}} + \sqrt {{\beta _{\emph{i}}}} {g_{{\rm{PS}},{\emph{i}}}}{g_{{\rm{SR}},{\emph{i}}}}{c_{\emph{i}}}} \right|}^2}}}{{{\sigma ^2}}}} \right)} \right]\! ,
\end{align}
where $W$ is the channel bandwidth. By comparing \eqref{2} with $W  {\tau _\emph{i}}  {\log _2}\left( {1 + \frac{{{P_p}{{\left| {{g_{\rm{PR}}}} \right|}^2}}}{{{\sigma ^2}}}} \right)$, which is the PT's rate derived in the absence of the $\text{SU}_{\emph{i}}$, it can be observed that the signal backscattered by the $\text{SU}_{\emph{i}}$ improves the PT's rate and its increased rate during BC can be formulated as
\begin{align}\label{3}
{\Delta _1} = \sum\limits_{\emph{i} = 1}^K {\left[ {R_p^{{\tau _\emph{i}}} - W{\tau _\emph{i}}{{\log}_2}\left( {1 + \frac{{{P_p}{{\left| {{g_{\rm{PR}}}} \right|}^2}}}{{{\sigma ^2}}}} \right)} \right]} .
\end{align}

The SIC technique is performed at the receiver to decode $\text{SU}_{\emph{i}}$'s signal ${c_\emph{i}}$. Assuming that the primary signal $s\left( n \right)$ can be perfectly removed from ${y_{b\emph{i}}}\left( n \right)$, we have
 \begin{align}\label{4}
\hat y_{b\emph{i}}\left( n \right) = \sqrt {{P_p}} \sqrt {{\beta _\emph{i}}} {c_\emph{i}}{g_{\rm{PS},{\emph{i}}}}{g_{\rm{SR},{\emph{i}}}}s(n) + w\left( n \right).
\end{align}

Since the symbol ${c_\emph{i}}$ is received in \emph{N} consecutive primary symbol periods, the primary signal $s\left( n \right)$ can be regarded as spread-spectrum code with length $\emph{N}$ for ${c_\emph{i}}$. By using the maximal ratio combing (MRC)\footnote{Perfect SIC and MRC are assumed to ensure fair comparison with related works and are consistent with assumptions in prior studies \cite{8907447,9866050,9193946}.}, the signal-to-noise ratio of ${c_\emph{i}}$ is increased by $\emph{N}$ times at the cost of symbol rate decreased by $\frac{1}{\emph{N}}$, and the rate of $\text{SU}_{\emph{i}}$ during ${\tau_\emph{i}}$ is expressed as
\begin{align}\label{5}
R_{{s}}^{{\tau _\emph{i}}} = \frac{1}{\emph{N}}W  {\tau _\emph{i}} {\log _2}\left( {1 + \frac{{\emph{N}{\beta _\emph{i}}{P_p}{{\left| {{g_{\rm{PS},\emph{i}}}{g_{\rm{SR},\emph{i}}}} \right|}^2}}}{{{\sigma ^2}}}} \right).
\end{align}

 Different from the traditional mutualistic SR where SUs are assumed to only harvest energy from the primary signal, in this work, we consider that each SU  harvests energy from the PT's signal as well as from signals of SUs for energy recycling. By doing so, each SU harvests more energy and achieves a higher rate than that of the traditional mutualistic SR. Thus, the energy collected by the $\text{SU}_{\emph{i}}$ within $\emph{T}_b$ can be expressed as
\begin{align}\label{6}
{H_{b\emph{i}}^{har}} &= \underbrace {\eta \left( {1 - {\beta _\emph{i}}} \right){P_p}\;{\left| {{g_{\rm{PS},\emph{i}}}} \right|^2}{\tau _\emph{i}}}_{\rm{the}\; \rm{first}\; \rm{term}} +\underbrace {\eta\; {P_p}\;{\left| {{g_{\rm{PS},\emph{i}}}} \right|^2}\left( {{\emph{T}_b} - {\tau _\emph{i}}} \right)}_{\rm{the}\; \rm{second}\; \rm{term}} \notag \\
&+\underbrace {\eta{P_p}\sum\nolimits_{j = 1,j \ne i}^{\emph{K}} {{{\left| {{g_{\rm{PS},\emph{j}}}} \right|}^2}{{\left| {{g_{\rm{SS},\emph{ji}}}} \right|}^2}}{\beta _\emph{j}} {\tau _\emph{j}}}_{\rm{the}\; \rm{third}\; \rm{term}},
\end{align}
where $\eta$ is the EH coefficient, the first term represents the energy collected from the primary signal
 during $\tau_\emph{i}$, the second term is the energy collected from the primary signal during the period that other $(\emph{K}-1)$ users perform BC, the third term denotes the energy harvested from the backscattered signal for energy recycling.

\subsection{Active Communication Phase}
During ${\emph{T}_a}$, PT and SUs transmit signal via uplink non-orthogonal multiple access (NOMA).
In what follows, we analyze the PT's and SU's rates under the fixed SIC ordering, yet, both the PT's and SU's rates under the dynamic SIC ordering will be examined in Section IV. In this phase, each SU takes turns to perform AC and continues to harvest energy. Specifically, when the $\text{SU}_{\emph{i}}$ transmits signal ${x_\emph{i}}\left( n \right)$ to the receiver during ${t_\emph{i}}$, others SUs harvest energy from the primary signal and the $\text{SU}_{\emph{i}}$'s active signal. From the information transmission's perspective, the received signal ${y_{a\emph{i}}}$ is represented as
\begin{align}\label{7}
{y_{a\emph{i}}}\left( n \right) = \underbrace {\sqrt {{P_p}} {g_{\rm{PR}}}s\left( n \right)}_{{\rm{PT}} \to {\rm{receiver\; link}}} + \underbrace {\sqrt {{P_{tr\emph{i}}}} {g_{\rm{SR},\emph{i}}}{x_\emph{i}}\left( n \right)}_{{\rm{S}}{{\rm{U}}_\emph{i}} \to {\rm{receiver\; link}}} + w\left( n \right),
\end{align}
where ${{P_{tr\emph{i}}}}$ is the transmit power of $\text{SU}_{\emph{i}}$.

Then the PT's rate during ${t_\emph{i}}$ is expressed as
\begin{align}\label{8}
R_{p1}^{{t_\emph{i}}} = W  {t_\emph{i}} {\log _2}\left( {1 + \frac{{{P_p} {{\left| {{g_{\rm{PR}}}} \right|}^2}}}{{{P_{tr\emph{i}}}{{\left| {{g_{\rm{SR},\emph{i}}}} \right|}^2} + {\sigma ^2}}}} \right).
\end{align}

Obviously, the PT's rate shown in \eqref{8} is strictly lower than the PT's rate derived without the presence of SU, i.e., $W  {t_{\emph{i}}}  {\log _2}\left( {1 + \frac{{{P_p}{{\left| {{g_{\rm{PR}}}} \right|}^2}}}{{{\sigma ^2}}}} \right)$, due to the interference ${{P_{tr\emph{i}}}{{\left| {{g_{\rm{SR},\emph{i}}}} \right|}^2}}$ introduced by the $\text{SU}_{\emph{i}}$. This indicates that within $t_\emph{i}$, the AC of the $\text{SU}_{\emph{i}}$ decreases the PT's rate. The decreased rate during AC can be given as

\begin{align}\label{9}
{\Delta _2} = \sum\limits_{\emph{i} = 1}^{\emph{K}} {W{t_\emph{i}}}& \left[ {{{\log }_2}\left( {1 + \frac{{{P_p}{{\left| {{g_{\rm{PR}}}} \right|}^2}}}{{{P_{tr\emph{i}}}{{\left| {{g_{\rm{SR},\emph{i}}}} \right|}^2} + {\sigma ^2}}}} \right)} \right.\notag \\
&\left. { - {{\log }_2}\left( {1 + \frac{{{P_p}{{\left| {{g_{\rm{PR}}}} \right|}^2}}}{{{\sigma ^2}}}} \right)} \right] .
\end{align}

Once the primary signal is decoded, we assume that it can be perfectly removed via SIC. Then  we have
\begin{align}\label{10}
{\hat y_{a\emph{i}}}(n) = \sqrt {{P_{tr\emph{i}}}} {g_{\rm{SR},\emph{i}}}{x_\emph{i}}\left( n \right) + w\left( n \right).
\end{align}

Using \eqref{10}, the rate of the $\text{SU}_{\emph{i}}$ within ${t_\emph{i}}$ can be written as
\begin{align}\label{11}
R_{{s_1}}^{{t_\emph{i}}} = W  {t_\emph{i}}  {\log _2}\left( {1 + \frac{{{P_{tr\emph{i}}} {{\left| {\rm{g_{{SR},\emph{i}}}} \right|}^2}}}{{{\sigma ^2}}}} \right).
\end{align}

From the EH's perspective, within $\emph{T}_a$, each SU harvests energy from signals of PT and SUs for energy recycling, then we have
\begin{align}\label{12}
{H_{a\emph{i}}^{har}}\! =\! \eta {P_p}{\left| {{g_{\rm{PS},\emph{i}}}} \right|^2}\!\!\left( {{\emph{T}_a}\! -\! {t_\emph{i}}} \right)\! + \eta \! \sum\limits_{\emph{j} = 1,\emph{j} \ne i}^\emph{K} \!{{P_{tr\emph{j}}}{{\left| {{g_{\rm{SS},\emph{ji}}}} \right|}^2}} {t_\emph{j}}.
\end{align}

\subsection{Energy Harvesting and Consumption of SUs}
Using \eqref{6} and \eqref{12}, we have obtained the harvested energy of $\text{SU}_{\emph{i}}$, i.e., ${H_{\emph{i}}^{har}} = H_{b\emph{i}}^{har} + H_{a\emph{i}}^{har}$. In what follows, we analyze the $\text{SU}_{\emph{i}}$'s energy consumption during BC and AC phases. Particularly, the $\text{SU}_{\emph{i}}$ consumes energy only in ${\tau _\emph{i}}$ and ${t_\emph{i}}$. During ${\tau _\emph{i}}$, the $\text{SU}_{\emph{i}}$ modulates its own information over the primary signal and does not need to generate radio frequency carrier. Thus, there is only circuit consumption, i.e., $H_{b\emph{i}}^{con} = {\varepsilon _b} {\tau _\emph{i}}$ with the static circuit power consumption of BC ${\varepsilon _b}$ \cite{9161012}. In ${t_\emph{i}}$, the $\text{SU}_{\emph{i}}$ adopts AC to transmit information, thus, its energy consumption includes transmit power and the circuit operation, i.e., $H_{a\emph{i}}^{con} ={P_{tr\emph{i}}}  {t_\emph{i}} + {\varepsilon _a}  {t_\emph{i}}$ with the AC's static circuit power consumption ${\varepsilon _a}$.

\section{Resource Allocation with Fixed SIC Ordering}
In this section, we formulate an optimization problem to maximize the SUs' total rate by jointly optimizing the SUs' reflection coefficients, the transmit power of each SU during AC, and the time allocation for each SU between BC and AC, while considering the fixed SIC ordering during AC. The problem is non-convex and challenging to solve. Accordingly, we use SCA and auxiliary variables to transform the original problem into a convex one. Lastly, a SCA-based iterative algorithm is proposed to solve it.

\subsection{Problem Formulation}
The goal of this design is to maximize the total rate of all SUs under the constraint of mutualistic relationship between the PT and SUs, and the energy-causality constraint of each SU. The optimization problem is formulated as follows:
\begin{align}\label{13}
&{{\bf{P}}_1:} \mathop {\max }\limits_{{\tau _\emph{i}},{t_\emph{i}},{\beta _\emph{i}},{P_{tr\emph{i}}},{\emph{T}_a},{\emph{T}_b}} \sum\nolimits_{\emph{i} = 1}^\emph{K} {\left( {R_{s}^{{\tau _\emph{i}}} + R_{{s_1}}^{{t_\emph{i}}}} \right)} \\
&{\;\;{\rm{s}}.{\rm{t}}.\; {\Delta _1} + {\Delta _2} \ge \Delta} ,\tag{13a}\label{13a}\\
&{\;\;\;\;\;\;\;H_{b\emph{i}}^{har} + H_{a\emph{i}}^{har} \ge H_{b\emph{i}}^{con} + H_{a\emph{i}}^{con},\forall \emph{i}},\tag{13b}\label{13b}\\
&{\;\;\;\;\;\;\;0 \le {\beta _\emph{i}} \le 1,\forall \emph{i}},\tag{13c}\label{13c}\\
&{\;\;\;\;\;\;\;{P_{tr\emph{i}}} \ge 0,\forall \emph{i}},\tag{13d}\label{13d}\\
&{\;\;\;\;\;\;\;{\tau _\emph{i}} \ge 0,{t_\emph{i}} \ge 0,\forall \emph{i}},\tag{13e}\label{13e}\\
&{\;\;\;\;\;\;\;\sum\nolimits_{\emph{i} = 1}^\emph{K} {{\tau _\emph{i}}}  \le {\emph{T}_b}},\tag{13f}\label{13f}\\
&{\;\;\;\;\;\;\;\sum\nolimits_{\emph{i} = 1}^\emph{K} {{t_\emph{i}}}  \le {\emph{T}_a}},\tag{13g}\label{13g}\\
&{\;\;\;\;\;\;\;{\emph{T}_a} + {\emph{T}_b} \le T},\tag{13h}\label{13h}
\end{align}
where $\Delta$ is a positive constant denoting the minimum rate gain\footnote{PT's rate gain is defined as the difference between the PT's rate when SUs are allowed to share spectrum resource with PT and that when the spectrum resource is only used by the PT. Hence,  PT's rate gain equals ${\Delta _1} + {\Delta _2}$.} of the PT. Constraint (13a) ensures that the sum of the increased rate and the decreased rate is not smaller than $\Delta $. This means that in our network, the PT's rate can still be improved by allowing SUs to perform HAPC, which is essential to ensure the mutualistic relationship. Constraint (13b) ensures that the energy consumed by $\text{SU}_{\emph{i}}$ for information transmission can be compensated by the harvested energy.
Constraints (13c) and (13d) limit the range of the reflection coefficient and the transmit power for each SU. Constraints (13f) and (13g) indicate that the total transmission time of each SU for BC  and AC cannot exceed the allocated time. Constraint (13h) ensures that the sum of the BC time of all SUs and the AC time of all SUs is smaller than $\emph{T}$.

\subsection{Problem Transformation and Iterative Algorithm}
${{\bf{P}}_1}$ is non-convex due to the presence of multiple coupled variables in the objective function and constraints. Specifically, ${\tau _\emph{i}}$ and ${\beta _\emph{i}}$ are coupled in $R_{{s}}^{{\tau _\emph{i}}}$, which appear in the objective function and in constraint (13a), as well as in constraint (13b). Besides,  $R_{{s_1}}^{{t_\emph{i}}}$ in the objective function, constraints (13a) and (13b) also have coupled variables with respect to $t_\emph{i}$ and $P_{tr\emph{i}}$. To address this issue, two steps are performed to decouple these optimization variables. Firstly, the SCA is employed to transform a non-convex function into a linear one. Secondly, auxiliary variables $q_\emph{i}$ and $z_\emph{i}$ are constructed to decouple ${\tau _\emph{i}}  {\beta _\emph{i}}$ and ${t_\emph{i}} {P_{tr\emph{i}}}$. The detailed procedure is stated as follows.

\begin{algorithm}[!t]
\setstretch{1}
\caption{SCA-Based Iterative Algorithm}
\label{alg:A}
\renewcommand{\algorithmicrequire}{\textbf{Input:}}
\renewcommand{\algorithmicensure}{\textbf{Output:}}
\begin{algorithmic}[1]
\REQUIRE  The location of the receiver, SUs, and PT, $P_p$, $\eta$, $\sigma ^2$, $\emph{T}$, $W$, ${\varepsilon _b}$, ${\varepsilon _a}$, ${\xi _\emph{i}}$, ${\alpha _\emph{i}}$
\ENSURE $\tau _\emph{i}^ * $, $t_i^ *$, $\beta _\emph{i}^*$, $\emph{T}_a^ * $, $\emph{T}_b^ * $, $P_{tr\emph{i}}^ * $
\STATE { Initialize ${P_{tr\emph{i}}^{(\emph{L})}},\forall \emph{i}$, iteration index $L = 0$, the maximum iterations $\emph{L}_{\max}$, and $\rm{Flag}=0$;}
\REPEAT
\STATE Solve ${\bf{P}}_3$  and obtain $\tau _\emph{i}^{\emph{(L)} }$, $t_\emph{i}^{\emph{(L)} }$, $q_\emph{i}^{\emph{(L)} }$, $z_\emph{i}^{\emph{(L)} }$, $\emph{T}_a^{\emph{(L)} }$, $\emph{T}_b^{\emph{(L)}}$, ${P_{tr\emph{i}}^\emph{(L)}}$;
\IF {the objective value (17) converges}
\STATE Set $\tau _\emph{i}^ * \! =\! \tau _\emph{i}^{\emph{(L)}}$, $t_i^ * \! = \! t_\emph{i}^{\emph{(L)} }$, $q_\emph{i}^ * \! = \! q_\emph{i}^{\emph{(L)} }$,
 $z_\emph{i}^ * \! =\! z_\emph{i}^{\emph{(L)}}$, $\emph{T}_a^ * \! =\! \emph{T}_a^{(L)}$, $\emph{T}_b^ * \! =\! \emph{T}_b^{\emph{(L)}}$, $\rm{Flag}=1$;
 \STATE Compute  $P_{tr\emph{i}}^ *  = {z_\emph{i}^{*} \mathord{\left/
 {\vphantom {z_\emph{i}^{*} t_\emph{i}^{* }}} \right.
 \kern-\nulldelimiterspace} t_\emph{i}^{* }}$, $\beta _\emph{i}^* = {{q_\emph{i}^{*}} \mathord{\left/
 {\vphantom {{q_\emph{i}^{*}} {\tau _\emph{i}^{*}}}} \right.
 \kern-\nulldelimiterspace} {\tau _\emph{i}^{*}}}$;
\ELSE
\STATE Update ${P_{tr\emph{i}}^{(\emph{L})}}$, $\emph{L}=\emph{L}+1$ and $\rm{Flag}=0$;
\ENDIF
\UNTIL{$\emph{L}=\emph{L}_{\max}$ or $\rm{Flag}=1$.}
\end{algorithmic}
\end{algorithm}

Let ${a_\emph{i}} = {\left| {{g_{\rm{PS},\emph{i}}}} \right|^2}$, $b = {\left| {{g_{\rm{PR}}}} \right|^2}$, ${d_\emph{i}} = {\left| {{g_{\rm{SR},\emph{i}}}} \right|^2}$ and ${f_\emph{ji}} = {\left| {{g_{\rm{SS},\emph{ji}}}} \right|^2}$. $R_{{p_1}}^{{t_\emph{i}}}=W  {t_\emph{i}} {\log _2}\left( {1 + \frac{{{P_{p }}  b}}{{{\sigma ^2} + {P_{tr\emph{i}}} {d_\emph{i}}}}} \right)$ is convex with respect to $P_{tri}$, which results in (13a) being a reverse-convex constraint \cite{razaviyayn2014successive}. We denote $R_{{p_1}}^{{t_\emph{i}}}$ as $f(P_{tri})$ for clarity. To transform constraint (13a) into a convex one, one way is to approximate $f(P_{tri})$ with its lower bound in an iterative manner as shown in \textbf{Lemma 1}.

\textbf{Lemma 1}: The strictly lower bound of $f(P_{tri})$ is $f\left( {P_{tr\emph{i}}^{(\emph{L})}} \right) + f'\left( {P_{tr\emph{i}}^{(\emph{L})}} \right)\left( {{P_{tr\emph{i}}} - P_{tr\emph{i}}^{(\emph{L})}} \right)$.

\emph{Proof.} The first derivative of $f\left( {P_{tr\emph{i}}} \right)$ with respect to $P_{tri}$ is
\begin{align}\label{14}
f'\left( {{P_{tr\emph{i}}}} \right)\! =\! \frac{{-W  {t_\emph{i}}  {P_{p }} b {d_\emph{i}}}}{{\ln 2\left( {{\sigma ^2}\! +\! {P_{tr\emph{i}}}{d_\emph{i}}\! +\! {P_{p }}b} \right)\left( {{\sigma ^2} \!+\! {P_{tr\emph{i}}}{d_\emph{i}}} \right)}},\forall \emph{i}. 
\end{align}
By taking the derivative of $f'\left( {{P_{tr\emph{i}}}} \right)$, the second derivative of $f(P_{tri})$ with respect to $P_{tri}$ is expressed as
$f''\left( {{P_{tri}}} \right) = \frac{{W{t_i}bd_i^2\left( {2{\sigma ^2} + 2{P_{tri}}{d_i} + {P_p}b} \right)}}{{\ln 2{{\left[ {\left( {{\sigma ^2} + {P_{tri}}{d_i} + {P_p}b} \right)\left( {{\sigma ^2} + {P_{tri}}{d_i}} \right)} \right]}^2}}}$.
It's obviously that $f(P_{tri})$ is convex due to $f''\left( {{P_{tri}}} \right) > 0$ \cite{boyd2004convex}. According to the principle of SCA \cite{razaviyayn2014successive}, the first order Taylor expansion of $f(P_{tri})$ serves as a strict lower bound and is represented as
\begin{align}\label{15}
f\left( {{P_{tr\emph{i}}}} \right) \ge f\left( {P_{tr\emph{i}}^{(\emph{L})}} \right) + f'\left( {P_{tr\emph{i}}^{(\emph{L})}} \right)\left( {{P_{tr\emph{i}}} - P_{tr\emph{i}}^{(\emph{L})}} \right), \forall \emph{i}, 
\end{align}
where $P_{tr\emph{i}}^{ (\emph{L}) }$ denotes the obtained value after \emph{L}-th iteration, and the equality in (16) only hold when ${P_{tr\emph{i}}} = P_{tr\emph{i}}^{ (\emph{L}) }$. \hfill {$\blacksquare $}

Substituting \eqref{15} into (13a) of ${{\bf{P}}_1}$, we have ${{\bf{P}}_2} $, as shown at the top of the next page.

\begin{figure*}
\begin{align}
{{\bf{P}}_2:}&\mathop {\max }\limits_{{\tau _\emph{i}},{t_\emph{i}},{\beta _\emph{i}},{P_{tr\emph{i}}},{\emph{T}_a},{\emph{T}_b}} \sum\limits_{\emph{i} = 1}^\emph{K} {\left( {\frac{{W {\tau _\emph{i}}}}{\emph{N}}{{\log }_2}\left( {1 + \frac{{\emph{N}{\beta _\emph{i}}{P_p} {a_\emph{i}}{d_\emph{i}}}}{{{\sigma ^2}}}} \right)} \right.}+ \left. {W  {t_\emph{i}}  {{\log }_2}\left( {1 + \frac{{{P_{tr\emph{i}}}{d_\emph{i}}}}{{{\sigma ^2}}}} \right)} \right)\label{16}\\
{\rm{s}}.{\rm{t}}.\notag \\
&\sum\limits_{\emph{i} = 1}^K \!{\left\{\! {\!W{\tau _\emph{i}}{{\mathbb{E}}_{{c_\emph{i}}}}\!\!\left[\! {{{\log }_2}\!\left(\! {1 \!+\! \frac{{{P_p}{{\left| {\sqrt b \! +\! \sqrt {{a_\emph{i}}{d_\emph{i}}{\beta _\emph{i}}} {c_\emph{i}}} \right|}^2}}}{{{\sigma ^2}}}}\! \right)}\! \right]\!\! +\! f\!\!\left( {\!P_{tr\emph{i}}^{(\!\emph{L}\!)}}\! \right)\! +\! f'\!\left( {\!P_{tr\emph{i}}^{(\!\emph{L}\!)}}\! \right)\!\!\left( \!{{\!P_{tr\emph{i}}}\! -\! P_{tr\emph{i}}^{(\!\emph{L}\!)}} \!\right)\! -\! W\left( {{\tau _\emph{i}}\! +\! {t_\emph{i}}} \right){{\log }_2}\!\left( {1\! + \!\frac{{{P_p}b}}{{{\sigma ^2}}}}\! \right)} \right\}}\ge {\Delta}, \tag{16a}\label{16a}\\
&\eta {P_p}{a_\emph{i}}\left( {1 \!-\! {\beta _\emph{i}}} \right){\tau _\emph{i}} + \eta {P_p}{a_\emph{i}}\left( {{\emph{T}_b} - {\tau _\emph{i}}} \right) + \eta {P_p}{a_\emph{i}}\left( {{\emph{T}_a}\! -\! {t_\emph{i}}} \right)\! + \!\eta {P_p}\sum\limits_{\emph{j} = 1,\emph{j} \ne \emph{i}}^K  {a_\emph{j}}{f_{\emph{ji}}}{\beta _\emph{j}}{\tau _\emph{j}}\! +\! \eta\! \sum\limits_{\emph{j} = 1,\emph{j} \ne \emph{i}}^K  {P_{tr\emph{j}}}{f_{\emph{ji}}}{t_\emph{j}}{\rm{ }} \ge {\varepsilon _b}  {\tau _\emph{i}}\! +\! {\varepsilon _a}  {t_\emph{i}} \!+\! {P_{tr\emph{i}}} {t_\emph{i}}, \forall \emph{i}, \tag{16b}\label{16b}\\
&(13\rm{c})-(13\rm{h}). \tag{16c}\label{16c}
\end{align}
\hrulefill
\end{figure*}

Although ${{\bf{P}}_2}$ is more tractable than ${{\bf{P}}_1}$, the objective function and constraints, i.e., (16a), (16b), still have several coupled variables such as ${\tau _\emph{i}}$ and ${\beta _\emph{i}}$, ${t_\emph{i}}$ and ${P_{tr\emph{i}}}$, which make the problem non-convex and require further transformation. Thereby, we introduce auxiliary variables ${q_\emph{i}} = {\tau _\emph{i}}  {\beta _\emph{i}}$ and ${z_\emph{i}} = {t_\emph{i}} {P_{tr\emph{i}}}$ to decouple the coupled variables. Then ${{\bf{P}}_2}$ is  transformed into ${{\bf{P}}_3}$, as shown
at the top of the next page.

\begin{figure*}
\begin{align}
{{\bf{P}}_3:}&\mathop {\max }\limits_{{\tau _\emph{i}},{t_\emph{i}},{q_\emph{i}},{z_\emph{i}},{\emph{T}_a},{\emph{T}_b}} \sum\limits_{\emph{i} = 1}^K {\left( {\frac{{W{\tau _\emph{i}}}}{N}{{\log }_2}\left( {1 + \frac{{N{q_\emph{i}}{P_p}{a_\emph{i}}{d_\emph{i}}}}{{{\sigma ^2}{\tau _\emph{i}}}}} \right)} \right.} \left. { + W{t_\emph{i}}{{\log }_2}\left( {1 + \frac{{{z_\emph{i}}{d_\emph{i}}}}{{{t_\emph{i}}{\sigma ^2}}}} \right)} \right)\label{17}\\
{\rm{s}}.{\rm{t}}.\notag \\
&\sum\limits_{\emph{i} = 1}^K \!{\left\{\! {W{\tau _\emph{i}}{{\mathbb{E}}_{{c_\emph{i}}}}\!\left[ {{{\log }_2}\left( {1\! + \frac{{{P_p}{{\left| {\sqrt b \! + \sqrt {{{{a_\emph{i}}{d_\emph{i}}{q_\emph{i}}} \mathord{\left/ {\vphantom {{{a_\emph{i}}{d_\emph{i}}{q_\emph{i}}} {{\tau _\emph{i}}}}} \right. \kern-\nulldelimiterspace} {{\tau _\emph{i}}}}} {c_\emph{i}}} \right|}^2}}}{{{\sigma ^2}}}} \right)}\! \right]\! -\! W\left( {{\tau _\emph{i}}\! + {t_\emph{i}}} \right){{\log }_2}\left( {1 + \frac{{{P_p}b}}{{{\sigma ^2}}}} \right) + W{t_\emph{i}}{{\log }_2}\left( {1 + \frac{{{P_p}b}}{{{\sigma ^2} + P_{tr\emph{i}}^{(\emph{L})}{d_\emph{i}}}}} \right)} \right.} \notag \\
&\;\;\;\;\;\;\;\;\;\left. { - \frac{{W{P_p}b{d_\emph{i}}\left( {{z_\emph{i}} - {t_\emph{i}}P_{tr\emph{i}}^{(\emph{L})}} \right)}}{{\ln 2\left( {{\sigma ^2} + P_{tr\emph{i}}^{(\emph{L})}{d_\emph{i}} + {P_p}b} \right)\left( {{\sigma ^2} + P_{tr\emph{i}}^{(\emph{L})}{d_\emph{i}}} \right)}}} \right\}\ge {\Delta},  \tag{17a}\label{17a}\\
&\eta {P_p} {a_\emph{i}}\left( {{\tau _\emph{i}}\! -\! {q_\emph{i}}} \right)\! +\! \eta {P_p} {a_\emph{i}} \left( {{\emph{T}_b}\!\! -\! {\tau _\emph{i}}} \right)\! +\! \eta {\!P_p} {a_\emph{i}} \left( {{\emph{T}_a}\! -\! {t_\emph{i}}} \right)+ \eta {P_p}\sum\nolimits_{\emph{j} = 1,\emph{j} \ne \emph{i}}^K {\!{a_\emph{j}}{f_{\emph{ji}}}{q _\emph{j}}}\!  +\! \eta \sum\nolimits_{\emph{j} = 1,\emph{j} \ne \emph{i}}^K {{\!f_{\emph{ji}}}} {z_\emph{j}} \ge {\varepsilon _b} {\tau _\emph{i}}\! +\! {\varepsilon _a} {t_\emph{i}}\!+\! {z_\emph{i}},\forall \emph{i}, \tag{17b}\label{17b}\\
&0 \le {q_\emph{i}} \le {\tau _\emph{i}},\forall \emph{i},\tag{17c}\label{17c}\\
&{z_\emph{i}} \ge 0,\forall \emph{i},\tag{17d}\label{17d}\\
&(13\rm{e})-(13\rm{h}). \tag{17e}\label{17e}
\end{align}
\hrulefill
\end{figure*}

It is observed that the objective function is a standard log-form concave function and the constraints from (17b)-(17e) are linear with respect to ${\tau _\emph{i}}$, ${t_\emph{i}}$, ${q_\emph{i}}$, ${z_\emph{i}}$, ${\emph{T}_a}$, ${\emph{T}_b}$. While for (17a), $W{\tau _\emph{i}}{{\mathbb{E}}_{{c_\emph{i}}}}\!\!\left[ {{{\!\log }_2}\!\left( {\!\!1 \!+ \! \frac{{{P_p}{{\left| {\sqrt b  + \sqrt {{{{a_\emph{i}}{d_\emph{i}}{q_\emph{i}}} \mathord{\left/ {\vphantom {{{a_\emph{i}}{d_\emph{i}}{q_\emph{i}}} {{\tau _\emph{i}}}}}\! \right. \kern-\nulldelimiterspace} {{\tau _\emph{i}}}}} {c_\emph{i}}} \right|}^2}}}{{{\sigma ^2}}}}\! \right)}\!\! \right]$ is a jointly concave function with respect to ${q_\emph{i}}$ and ${\tau _\emph{i}}$, \emph{as proven in Appendix A}, and the rest of (17a) is a linear function. Therefore, ${{\bf{P}}_3} $ is jointly convex with respect to all the  variables.
Based on the above transformations, we propose a SCA-based iterative algorithm in Algorithm \ 1 to solve ${{\bf{P}}_3} $. 

\subsection{Insights}
To obtain useful insights, we provide the following Propositions.

\textbf{Proposition 1.} For maximizing the total rate of SUs,  at least one equality holds in constraints (13a) and (13b).

\emph{Proof.} Please see Appendix B. \hfill {$\blacksquare $}

\emph{Remark 1:} Proposition 1 reveals the following two interesting facts. First, this Proposition indicates that there exists a possibility that not all SUs use up the harvested energy when ${{\bf{P}}_1}$ is optimally solved, and this seems contradictory to intuition. In previous works on rate maximization for wireless powered active/passive communications \cite{8571319,8340034} or wireless powered HAPC \cite{9161012,7981380,9534878}, all nodes should consume up the harvested energy.
However, in our considered network, it is possible that not all SUs need to use up their energy to maximize the transmission rate. This is due to the presence of the PT's rate gain constraint, $\Delta_1 + \Delta_2 \ge \Delta$, which couples the SUs' transmission strategies with the PT's performance. While using more energy can increase SU's transmission rate, it also causes stronger interference to the PT, thereby reducing $\Delta_2$. If $\Delta_2$ becomes too small, constraint (13a) may no longer be satisfied, making the solution infeasible. Thus, retaining part of the harvested energy may be optimal to balance SUs' transmission rate and PT's rate constraints.
 Second, in traditional mutualistic SR, SUs only perform BC and the BC will increase the PT's rate. In other words, the rate of PT is always increasing with the SUs' rates \cite{9866050}, and is not related to the it's minimum rate gain $\Delta$. However, in our considered mutualistic SR with HAPC, the PT's rate gain is increased during passive BC phase and decreased during AC phase. Specifically, when the channel conditions between PT and SUs become better, the $\text{SU}_{\emph{i}}$ harvests more energy and adopts a larger $P_{tr\emph{i}}$ to increase its transmission rate, which will decrease the left-hand side of constraint (13a), i.e., ${\Delta _1}+{\Delta _2}$, and then may go to $\Delta$ before the equality in (13b) holds.
In this case, the equality in (13a) holds. This indicates that there exists a tradeoff between the PT's rate gain and the total rate of SUs.

\textbf{Proposition 2.} The equality in (13h) holds when ${\bf{P}}_1$ is optimally solved.

\emph{Proof.} Please see Appendix C. \hfill {$\blacksquare $}

\emph{Remark 2:}  Proposition 2 indicates that, in our considered network, using up the total available time is an optimal choice. Actually, if the the total available time is not used up, increasing $\emph{T}_b$ or $\emph{T}_a$ can achieve a higher total rate of all SUs at a given PT's minimum rate gain. When $\emph{T}_b$ is increased, SUs not only contribute more multipath components to PT, but also harvest more energy and increase the SUs' rates. In such a case, both the SUs' rates and PT's rate gain increase, allowing more AC time for SUs to boost their total rates.

\section{Resource Allocation with Dynamic SIC Ordering}
In the proposed mutualistic SR with HAPC, SUs and the PT adopt uplink NOMA for information transmission during the AC phase, aiming to improve spectral and temporal efficiency. Under this setting, if PT's signal is decoded first during AC phase (as described in Section III), the SU introduces co-channel interference to the PT, which decreases PT's rate but increases SU's rate. Such a SIC ordering may not affect the total rate of all SUs when the PT's minimum rate gain $\Delta$ is small. However, if $\Delta$ is large, satisfying the PT's rate requirement may require reducing or even eliminating the AC time allocated to SUs, resulting in a significantly reduced  SUs' total rate.  This reveals a tradeoff between the rates of PT and SUs. To better explore this tradeoff and further enhance SUs' performance, we consider dynamic SIC ordering at the receiver during AC phase, where the receiver can choose to decode either the PT's or the SU's signal first.
By jointly optimizing the resources considered in Section III and the SIC ordering, the total rate of SUs can be further improved. Towards this end, we formulate a MIP optimization problem to maximize the total rate of SUs and solve it using a BCD-based iterative algorithm.

\subsection{Problem Formulation}
Since the dynamic SIC ordering is only included in the AC phase\footnote{During BC phase, if the $\text{SU}_{\emph{i}}$'s signal is decoded first, it will experience strong interference from the PT, resulting in a considerably lower rate compared to the one for decoding the PT's signal first, while the PT's rate remains the same in both SIC ordering schemes. Therefore, we consider decoding PT's signal first during BC phase.}, the analysis of the BC phase is the same as in Section II. As a result, here we only need to rewrite the rate expressions of PT and SUs during AC phase. In ${t_{\emph{i}}}$, $\text{SU}_{\emph{i}}$ performs AC and shares the same spectrum with the PT, forming an uplink NOMA. Then the receiver has two choices for SIC ordering. One is to decode PT's signal first and then to decode $\text{SU}_{\emph{i}}$'s signal after using SIC to remove PT's signal from the received signal. The other is to reverse the above SIC ordering. The above two types of SIC ordering result in different rates for the PT and $\text{SU}_{\emph{i}}$, as analyzed below.
\subsubsection{Decode PT's Signal First} In this case, the SIC ordering is the same as that considered in Section II-B. Thus PT's and $\text{SU}_{\emph{i}}$'s rates can be calculated as \eqref{8} and \eqref{11}, respectively, as shown in Section II-B.

\subsubsection{Decode $\text{SU}_{\emph{i}}$'s Signal First} In this case, the receiver first decodes $\text{SU}_{\emph{i}}$'s signal ${x_\emph{i}}\left( n \right)$ by treating PT's signal $s\left( n \right)$ as interference, and then decodes PT's signal after using SIC to remove $\text{SU}_{\emph{i}}$'s signal from the received signal. Then, within ${t_{\emph{i}}}$, the rates of $\text{SU}_{\emph{i}}$ and PT are expressed as
\begin{align}\label{19}
R_{s2}^{{t_\emph{i}}} = W{t_\emph{i}}{\log _2}\left( {1 + \frac{{{P_{tr\emph{i}}}{d_\emph{i}}}}{{{P_p}b + {\sigma ^2}}}} \right),
\end{align}
\begin{align}\label{20}
R_{p2}^{{t_\emph{i}}} = W{t_\emph{i}}{\log _2}\left( {1 + \frac{{{P_p}{b}}}{{{\sigma ^2}}}} \right).
\end{align}

It can be observed from \eqref{8} and \eqref{20}, \eqref{11} and \eqref{19}  that decoding $\text{SU}_{\emph{i}}$'s (or PT's signal) first results in a high (or small) PT's rate but a small (or high) $\text{SU}_{\emph{i}}$'s rate, indicating that there are different tradeoffs between two types of SIC ordering in terms of PT's rate and $\text{SU}_{\emph{i}}$'s rate during AC phase. By optimizing the aforementioned tradeoff, the rates of the SUs can be improved, which motivates us to adopt a dynamic SIC ordering strategy that jointly considers both SIC ordering schemes. With the dynamic SIC ordering, the rates of $\text{SU}_{\emph{i}}$ and PT within $t_\emph{i}$ can be written as
\begin{align}\label{201}
 { R_s^{{t_\emph{i}}}= {\alpha _{b\emph{i}}} R_{s1}^{{t_\emph{i}}} + {\alpha _{a\emph{i}}}R_{s2}^{{t_\emph{i}}}},
\end{align}
\begin{align}\label{202}
 { R_p^{{t_\emph{i}}}= {\alpha _{b\emph{i}}} R_{p1}^{{t_\emph{i}}} + {\alpha _{a\emph{i}}}R_{p2}^{{t_\emph{i}}}},
\end{align}
where ${\alpha _{b\emph{i}}}\in \left\{ {0,1} \right\}$, ${\alpha _{a\emph{i}}}\in \left\{ {0,1} \right\}$, and above two terms satisfy ${\alpha _{a\emph{i}}} + {\alpha _{b\emph{i}}} \le 1$. Specifically, the case with ${\alpha _{b\emph{i}}}=1$ and ${\alpha _{a\emph{i}}}=0$ corresponds to the SIC ordering where the PT's signal is decoded first; Otherwise, the receiver decodes $\text{SU}_{\emph{i}}$'s signal first.
Although dynamic SIC introduces some computational complexity, there are only two SIC orderings per time slot, keeping the complexity low. In addition, the SIC ordering remains stable over time, and prior works on NOMA have shown the practical feasibility of dynamic SIC \cite{9151196, 7959539}.

Then the optimization problem for maximizing the total rate of SUs is formulated as follows
\begin{align}
{{\bf{P}}_4:}
&\mathop {\max }\limits_{{\tau _\emph{i}},{t_\emph{i}},{\beta _\emph{i}},{P_{tr\emph{i}}},{\emph{T}_a},{\emph{T}_b},{\alpha _{b\emph{i}}},{\alpha _{a\emph{i}}}}  \!\!{\rm{ }}\sum\limits_{\emph{i} = 1}^K {\!\!\left( \!{ R_s^{{\tau _\emph{i}}}\! +\! {\alpha _{b\emph{i}}} R_{s1}^{{t_\emph{i}}}\! +\! {\alpha _{a\emph{i}}}R_{s2}^{{t_\emph{i}}}} \right)} \label{22}\\ \notag
{\rm{s}}.{\rm{t}}.
&\sum\limits_{\emph{i} = 1}^K {\left( { R_p^{{\tau _\emph{i}}} + {\alpha _{b\emph{i}}} R_{p1}^{{t_\emph{i}}} + {\alpha _{a\emph{i}}} R_{p2}^{{t_\emph{i}}}} - {R_0} \right)}  \ge \Delta ,\tag{22a}\label{22a}\\
\;\;\;\;\;&{\alpha _{a\emph{i}}},{\alpha _{b\emph{i}}} \in \left\{ {0,1} \right\},\forall \emph{i}{\rm{ }},\tag{22b}\label{22c}\\
\;\;\;\;\;&{\alpha _{a\emph{i}}} + {\alpha _{b\emph{i}}} \le 1,\forall \emph{i}{\rm{ }},\tag{22c}\label{22d}\\
&{\eqref{13b}-\eqref{13h}},\tag{22d}\label{22g}
\end{align}
where ${R_0}=W\left( {{\tau _{\emph{i}}} + {t_{\emph{i}}}} \right){\log _2}\left( {1 + {{{P_{p}}{d_{\emph{i}}}} \mathord{\left/
 {\vphantom {{{P_{tr{\emph{i}}}}{d_{\emph{i}}}} {{\sigma ^2}}}} \right.
 \kern-\nulldelimiterspace} {{\sigma ^2}}}} \right)$, constraint \eqref{22a} ensures that PT's rate gain is not smaller than $\Delta$, constraint \eqref{22c} and constraint \eqref{22d} indicate that the receiver can only adopt one type of SIC ordering within $t_{\emph{i}}$.
\subsection{Problem Transformation and Iterative Algorithm}
${\bf{P}}_4$ is more challenging to solve compared to ${\bf{P}}_1$. On the one hand, ${\bf{P}}_4$ has to optimize  additional integer variables that represent different types of SIC ordering, i.e.,  ${\alpha _{b\emph{i}}}$ and ${\alpha _{a\emph{i}}}$, and thereby becomes a MIP problem. On the other hand, except for coupled optimization variables that appear in ${\bf{P}}_1$, the integer optimization variable ${\alpha _{b\emph{i}}}$ and ${\alpha _{a\emph{i}}}$ are also coupled with ${{\tau _\emph{i}}}$, ${{t_\emph{i}}}$, ${{\beta _\emph{i}}}$ and ${{P_{tr\emph{i}}}}$ in ${\bf{P}}_4$. To tackle these non-convexities, we adopt the BCD method to decompose ${{\bf{P}}_4}$ into two sub-problems. In the first sub-problem, the variables ${{\tau _\emph{i}}}$, ${{t_\emph{i}}}$, ${{\beta _\emph{i}}}$, ${\emph{T}_a}$, ${\emph{T}_b}$, and ${{P_{tr\emph{i}}}}$ are jointly optimized for given SIC ordering, i.e., $\alpha _{b\emph{i}}$ and $\alpha _{b\emph{i}}$. In the second sub-problem, the SIC ordering is optimized based on the results obtained in the first sub-problem.

\subsubsection{Jointly Optimizing ${{\tau _\emph{i}}}$, ${{t_\emph{i}}}$, ${{\beta _\emph{i}}}$, ${T_a}$, ${T_b}$ , and ${{P_{tr\emph{i}}}}$}
For given $\alpha _{b\emph{i}}$ and $\alpha _{b\emph{i}}$, the first sub-problem ${{\bf{P}}_{4.1}}$ is formulated with the objective of maximizing the SUs' total rate, which is shown at the top of the next page.
\begin{figure*}
\begin{align}
{{\bf{P}}_{4.1}:} &\mathop {\max }\limits_{{\tau _\emph{i}},{t_\emph{i}},{\beta _\emph{i}},{\emph{T}_a},{\emph{T}_b},{P_{tr\emph{i}}}}  {\rm{ }}\!\sum\limits_{\emph{i} = 1}^K {\!\left( \!{\frac{1}{N}W{\tau _\emph{i}}{{\log }_2}\!\left( {1 + \frac{{N{\beta _\emph{i}}{P_p}{a_\emph{i}}{d_\emph{i}}}}{{{\sigma ^2}}}} \right) + {\alpha _{bi}}W{t_\emph{i}}{{\log }_2}\left( {1 + \frac{{{P_{tr\emph{i}}}{d_\emph{i}}}}{{{\sigma ^2}}}} \right) + {\alpha _{a\emph{i}}}W{t_\emph{i}}{{\log }_2}\left( {1 + \frac{{{P_{tr\emph{i}}}{d_\emph{i}}}}{{{P_p}{b} + {\sigma ^2}}}} \right)} \right)}  \label{23}\\
{\rm{s}}.{\rm{t}}.\notag \\
&\sum\limits_{\emph{i} = 1}^K {\left\{ {W{\tau _\emph{i}}{{\mathbb{E}}_{{c_\emph{i}}}}\left[ {{{\log }_2}\left( {1 + {{{P_p}{{\left| {\sqrt b  + \sqrt {{\beta _\emph{i}}{a_\emph{i}}{d_\emph{i}}} {c_\emph{i}}} \right|}^2}} \mathord{\left/
 {\vphantom {{{P_p}{{\left| {\sqrt b  + \sqrt {{\beta _i}{a_\emph{i}}{d_\emph{i}}} {c_\emph{i}}} \right|}^2}} {{\sigma ^2}}}} \right.
 \kern-\nulldelimiterspace} {{\sigma ^2}}}} \right)} \right] + {\alpha _{b\emph{i}}}W{t_\emph{i}}{{\log }_2}\left( {1 + \frac{{{P_p}b}}{{{P_{tr\emph{i}}}{d_\emph{i}} + {\sigma ^2}}}} \right) + {\alpha _{a\emph{i}}}W{t_\emph{i}}{{\log }_2}\left( {1 + \frac{{{P_p}b}}{{{\sigma ^2}}}} \right)} \right.} \notag\\
&\left. { - W\left( {{\tau _\emph{i}} + {t_\emph{i}}} \right){{\log }_2}\left( {1 + \frac{{{P_p}b}}{{{\sigma ^2}}}} \right)} \right\} > \Delta, \tag{23a}\label{23a}\\
&(13\rm{b})-(13\rm{h}).\tag{23b}\label{23c}
\end{align}
\hrulefill
\end{figure*}

 In ${{\bf{P}}_{4.1}}$, ${\tau _\emph{i}}$ and ${\beta _\emph{i}}$ are coupled in the objective function and constraints \eqref{23a} and \eqref{13b}. Besides, ${t_\emph{i}}$ and ${P_{tr\emph{i}}}$ are coupled both in the objective function and constraint \eqref{13b}, and the term ${W{t_\emph{i}}{{\log }_2}\left( {1 + \frac{{{P_p}b}}{{{P_{tr\emph{i}}}{d_\emph{i}} + {\sigma ^2}}}} \right)}$ in constraint \eqref{23a} is also non-convex. To address the above nonconvexities, we firstly use SCA to approximate ${W{t_\emph{i}}{{\log }_2}\left( {1 + \frac{{{P_p}b}}{{{P_{tr\emph{i}}}{d_\emph{i}} + {\sigma ^2}}}} \right)}$ as a linear function, where the first-order Taylor expansion with respect to ${P_{tr\emph{i}}}$ is similar to \eqref{15}. Then, to tackle the coupling issues, we introduce auxiliary variables ${\mu _{\emph{i}}}$ and ${\chi _\emph{i}}$, where ${\mu _{\emph{i}}} = {\tau _\emph{i}}{\beta _\emph{i}}$, ${\chi _\emph{i}} = {P_{tr\emph{i}}}{t_\emph{i}}$. The transformed optimization problem is denoted as  ${{\bf{P}}_{4.11}}$, as shown at the top of the next page. Please note that $P_{tr\emph{i}}^{ (\emph{u}) }$ represents the obtained value after \emph{u}-th iteration.
 \begin{figure*}
\begin{align}
{{\bf{P}}_{4.11}:}&\mathop {\max }\limits_{{\tau _\emph{i}},{t_\emph{i}},{\mu _\emph{i}},{\chi _\emph{i}},{\emph{T}_a},{\emph{T}_b}} \!\!\sum\limits_{\emph{i} = 1}^K {\left( {\!\frac{W{\tau _\emph{i}}}{N}{{\log }_2}\left( {1 + \frac{{N{\mu _i}{P_p}{a_\emph{i}}{d_\emph{i}}}}{{{\tau _\emph{i}}{\sigma ^2}}}} \right) + {\alpha _{b\emph{i}}}W{t_\emph{i}}{{\log }_2}\left( {1 + \frac{{{\chi _\emph{i}}{d_\emph{i}}}}{{{t_\emph{i}}{\sigma ^2}}}} \right) + {\alpha _{a\emph{i}}}W{t_\emph{i}}{{\log }_2}\left( {1 + \frac{{{\chi _\emph{i}}{d_\emph{i}}}}{{{t_\emph{i}}\left( {{P_p}b + \!{\sigma ^2}} \right)}}}\! \right)}\! \right)}\label{24}  \\
{\rm{s}}.{\rm{t}}.\notag \\
&\!\!\sum\limits_{\emph{i} = 1}^K {\left\{ {W{\tau _\emph{i}}{{\mathbb{E}}_{{c_\emph{i}}}}\left[ {{{\log }_2}\left( {1 + {{{P_p}{{\left| {\sqrt b  + \sqrt {{{{a_\emph{i}}{d_\emph{i}}{u_\emph{i}}} \mathord{\left/
 {\vphantom {{{a_\emph{i}}{d_\emph{i}}{u_\emph{i}}} {{\tau _\emph{i}}}}} \right.
 \kern-\nulldelimiterspace} {{\tau _\emph{i}}}}} {c_\emph{i}}} \right|}^2}} \mathord{\left/
 {\vphantom {{{P_p}{{\left| {\sqrt b  + \sqrt {{{{a_\emph{i}}{d_\emph{i}}{u_\emph{i}}} \mathord{\left/
 {\vphantom {{{a_\emph{i}}{d_\emph{i}}{u_\emph{i}}} {{\tau _\emph{i}}}}} \right.
 \kern-\nulldelimiterspace} {{\tau _\emph{i}}}}} {c_\emph{i}}} \right|}^2}} {{\sigma ^2}}}} \right.
 \kern-\nulldelimiterspace} {{\sigma ^2}}}} \right)} \right] + {\alpha _{a\emph{i}}}W{t_\emph{i}}{{\log }_2}\left( {1 + \frac{{{P_p}b}}{{{\sigma ^2}}}} \right) - W\left( {{\tau _i} + {t_\emph{i}}} \right){{\log }_2}\left( {1 + \frac{{{P_p}b}}{{{\sigma ^2}}}} \right)} \right.} \notag\\
&+{\alpha _{b\emph{i}}}W{t_\emph{i}}{\log _2}\left( {1 + \frac{{{P_p}b}}{{P_{tr\emph{i}}^{(\emph{u})}{d_\emph{i}} + {\sigma ^2}}}} \right)\left. { - \frac{{{\alpha _{b\emph{i}}}W{P_p}b{d_\emph{i}}\left( {{\chi _\emph{i}} - {t_\emph{i}}P_{tr\emph{i}}^{(\emph{u})}} \right)}}{{\ln 2\left( {{\sigma ^2} + P_{tr\emph{i}}^{(\emph{u})}{d_\emph{i}} + {P_p}b} \right)\left( {{\sigma ^2} + P_{tr\emph{i}}^{(\emph{u})}{d_\emph{i}}} \right)}}} \right\}> \Delta, \tag{24a}\label{24a}\\
&\eta\! \left( {{\!\tau _\emph{i}} - {\mu _\emph{i}}}\! \right)\!{P_p}{a_\emph{i}} \!+\eta {P_p}{a_\emph{i}}\left( {{\emph{T}_b} - {\tau _\emph{i}}} \right) +\eta {P_p}{a_\emph{i}}\left( {{\emph{T}_a} - {t_\emph{i}}} \right) +\! \eta {P_p}\!\!\sum\limits_{\emph{j} = 1,\emph{j} \ne \emph{i}}^K {\!\!{a_\emph{j}}{f_{\emph{ji}}}} {\mu _\emph{j}}\! +\! \eta \!\!\sum\limits_{\emph{j} = 1,\emph{j} \ne \emph{i}}^K {{f_{\emph{ji}}}} {\chi _\emph{j}} \ge {\varepsilon _b}{\tau _\emph{i}} + {\varepsilon _a}{t_\emph{i}} + {\chi _\emph{i}},\forall \emph{i}, \tag{24b}\label{24b}\\
&0 \le {\mu _\emph{i}} \le {\tau _\emph{i}},\forall \emph{i},\tag{24c}\label{24c}\\
&{\chi _\emph{i}} \ge 0,\forall \emph{i},\tag{24d}\label{24d}\\
&(13\rm{e})-(13\rm{h}), \forall \emph{i}. \tag{24e}\label{24e}
\end{align}
\hrulefill
\end{figure*}

It is observed from ${\bf{P}}_{4.11}$ that the objective function is a standard log-form concave function and constraints \eqref{24b}-\eqref{24e} are linear with respect to ${\tau _\emph{i}}$, ${t_\emph{i}}$, ${\mu _\emph{i}}$ and ${\chi _\emph{i}}$. The convexity of ${W\!{\tau _\emph{i}}{{\mathbb{E}}_{{c_\emph{i}}}}\!\!\left[ {{{\!\log }_2}\!\left( {\!\!1 \!\!+\! \frac{{{P_p}{{\left| {\!\sqrt b +\! \sqrt {{{{a_\emph{i}}{d_\emph{i}}{u_\emph{i}}} \mathord{\left/{\vphantom {{{a_\emph{i}}{d_\emph{i}}{u_\emph{i}}} {{\tau _\emph{i}}}}} \right. \kern-\nulldelimiterspace} {{\tau _\emph{i}}}}} {c_\emph{i}}}
\right|}^2}}}{{{\sigma ^2}}}}\!\! \right)}\!\! \right]}$ is the same as $W{\tau _\emph{i}}{{\mathbb{E}}_{{c_\emph{i}}}}\left[ {{{\log }_2}\left( {1 + \frac{{{P_p}{{\left| {\sqrt b  + \sqrt {{{{a_\emph{i}}{d_\emph{i}}{q_\emph{i}}} \mathord{\left/ {\vphantom {{{a_\emph{i}}{d_\emph{i}}{q_\emph{i}}} {{\tau _\emph{i}}}}} \right. \kern-\nulldelimiterspace} {{\tau _\emph{i}}}}} {c_\emph{i}}} \right|}^2}}}{{{\sigma ^2}}}} \right)} \right]$ in \eqref{17a} due to the same form of expression, and the rest expressions in constraint \eqref{24a} are linear with respect to ${\chi _\emph{i}}$ and ${t_\emph{i}}$. Thus, ${{\bf{P}}_{4.11}}$ is convex and can be solved by \emph{CVX} tools.

\subsubsection{Optimizing SIC ordering}
For given ${{\tau _\emph{i}}}$, ${{t_\emph{i}}}$, ${{\beta _\emph{i}}}$, ${\emph{T}_a}$, ${\emph{T}_b}$, and ${{P_{tr\emph{i}}}}$, the second sub-problem with the goal of maximizing the SUs' total transmission rate is expressed as
\begin{align}
{{\bf{P}}_{4.2}:}&\mathop {\max }\limits_{{\alpha _{a\emph{i}}},{\alpha _{b\emph{i}}}} {\rm{ }}\sum\limits_{\emph{i} = 1}^K {\left( { R{{_s^{{\tau _\emph{i}}}}^\prime } + {\alpha _{b\emph{i}}} R{{_{s1}^{{t_\emph{i}}}}^\prime } + {\alpha _{a\emph{i}}}R{{_{s2}^{{t_\emph{i}}}}^\prime }} \right)} \label{25}\\
{\rm{s}}.{\rm{t}}.\notag
&{\rm{ }}\sum\limits_{\emph{i} = 1}^K {\left\{ {R{{_p^{{\tau _\emph{i}}}}^\prime } + {\alpha _{b\emph{i}}} R{{_{p1}^{{t_\emph{i}}}}^\prime } + {\alpha _{a\emph{i}}}R{{_{p2}^{{t_\emph{i}}}}^\prime }}- {R_0} \right\}}   \ge \Delta ,  \tag{25a}\label{25a}\\
&\eqref{22c}, \eqref{22d}, \tag{25b}\label{25b}
\end{align}
where $ R{{_s^{{\tau _\emph{i}}}}^\prime} = {{W{\tau _\emph{i}}{{\log }_2}\left( {1 + \frac{{N{\mu _\emph{i}}{P_p}{a_\emph{i}}{d_\emph{i}}}}{{{\tau _\emph{i}}{\sigma ^2}}}} \right)} \mathord{\left/
 {\vphantom {{W{\tau _\emph{i}}{{\log }_2}\left( {1 + \frac{{N{\mu _\emph{i}}{P_p}{a_\emph{i}}{d_\emph{i}}}}{{{\tau _\emph{i}}{\sigma ^2}}}} \right)} N}} \right.
 \kern-\nulldelimiterspace} N}$, $ R{{_{s1}^{{t_\emph{i}}}}^\prime } = W{t_\emph{i}}{\log _2}\left( {1 + \frac{{{\chi _\emph{i}}{d_\emph{i}}}}{{{t_\emph{i}}{\sigma ^2}}}} \right)$, $R{{_{p2}^{{t_\emph{i}}}}^\prime } = W{t_\emph{i}}{\log _2}\left( {1 + \frac{{{P_p}b}}{{{\sigma ^2}}}} \right)$, $ R{{_p^{{\tau _\emph{i}}}}^\prime }= W{\tau _\emph{i}}{{\mathbb{E}}_{{c_\emph{i}}}}\left[ {{{\log }_2}\left( {1 + \frac{{{P_p}{{\left| {\sqrt b  + \sqrt {{{{a_\emph{i}}{d_\emph{i}}{u_\emph{i}}} \mathord{\left/
 {\vphantom {{{a_\emph{i}}{d_\emph{i}}{u_\emph{i}}} {{\tau _\emph{i}}}}} \right. \kern-\nulldelimiterspace} {{\tau _\emph{i}}}}} {c_\emph{i}}} \right|}^2}}}{{{\sigma ^2}}}} \right)} \right]$,  $R{{_{s2}^{{t_\emph{i}}}}^\prime }= W{t_\emph{i}}{\log _2}\left( {1 + \frac{{{\chi _\emph{i}}{d_\emph{i}}}}{{{t_\emph{i}}\left( {{P_p}b + {\sigma ^2}} \right)}}} \right)$and $R{_{p1}^{{t_\emph{i}}} }^\prime = W{t_\emph{i}}{\log _2}\left( {1 + \frac{{{P_p}b}}{{P_{tr\emph{i}}^{(\emph{u})}{d_\emph{i}} + {\sigma ^2}}}} \right) - \frac{{W{P_p}b{d_\emph{i}}\left( {{\chi _\emph{i}} - {t_\emph{i}}P_{tr\emph{i}}^{(\emph{u})}} \right)}}{{\ln 2\left( {{\sigma ^2} + P_{tr\emph{i}}^{(\emph{u})}{d_\emph{i}} + {P_p}b} \right)\left( {{\sigma ^2} + P_{tr\emph{i}}^{(\emph{u})}{d_\emph{i}}} \right)}}$.

\begin{algorithm}[!t]
\setstretch{1}
\caption{BCD-based Iterative Algorithm }
\label{alg:A}
\renewcommand{\algorithmicrequire}{\textbf{Input:}}
\renewcommand{\algorithmicensure}{\textbf{Output:}}
\begin{algorithmic}[1]
\REQUIRE  The location of the receiver, SUs, and PT, $P_p$, $\eta$, $\sigma ^2$, $\emph{T}$, $W$, ${\varepsilon _b}$, ${\varepsilon _a}$, ${\xi _\emph{i}}$, ${\alpha _\emph{i}}$
\ENSURE $\tau _\emph{i}^ * $, $t_\emph{i}^ * $, $\beta _\emph{i}^ * $, $P_{tr\emph{i}}^ * $, $\emph{T}_a^ *$ , $\emph{T}_b^ *$, $\alpha _{b\emph{i}}^ * $, $\alpha _{a\emph{i}}^ * $
\REPEAT
\STATE {Initialize the SIC ordering ${{\alpha _{b\emph{i}}^{(\gamma )}}}$, ${{\alpha _{a\emph{i}}^{(\gamma)}}}$, the iteration index $\gamma=0$;}
\REPEAT
\STATE {Initialize $\text{SU}_{\emph{i}}$'s transmit power $P_{tr\emph{i}}^{(\emph{u})}$, $\forall \emph{i}$, the iteration index $\emph{u}=0$;}
\STATE {Obtain $\left\{ {\tau _\emph{i}^ {\emph{u}} ,t_\emph{i}^ {\emph{u}},{\mu _\emph{i}^ {\emph{u}} },{\chi _\emph{i}^ {\emph{u}}, \emph{T}_a^ {\emph{u}} ,\emph{T}_b^ {\emph{u}} }} \right\}$ by solving ${\bf{P}}_{4.11}$;}
\STATE {Update $P_{tr\emph{i}}^{(u)}$ ;}
\UNTIL{the objective function \eqref{24} converges in the ${\bf{P}}_{4.11}$;}
\STATE {Obtain $\left\{ {\tau _\emph{i}^ {*'} ,t_\emph{i}^ {*'},{\mu _\emph{i}^ {*'} },{\chi _\emph{i}^ {*'}, \emph{T}_a^ {*'} ,\emph{T}_b^ {*'} }} \right\}$; }
\STATE {Compute $\beta _\emph{i}^{*'} = {{\mu _\emph{i}^ {*'} } \mathord{\left/
 {\vphantom {{\mu _\emph{i}^ {*'} } {\tau _\emph{i}^ {*'} }}} \right.
 \kern-\nulldelimiterspace} {\tau _\emph{i}^ {*'} }}$, $P_{tr\emph{i}}^* = {{\chi _\emph{i}^ {*'} } \mathord{\left/
 {\vphantom {{\chi _\emph{i}^ {*'} } {t_\emph{i}^ {*'} }}} \right.
 \kern-\nulldelimiterspace} {t_\emph{i}^ {*'} }}$ };
\REPEAT
\STATE {Initialize the SIC ordering $\alpha _{b\emph{i}}^{(\emph{r})}$, $\alpha _{a\emph{i}}^{(\emph{r})}$, the iteration index $\emph{r}=0$;}
\STATE {Obtain $\alpha _{b\emph{i}}^{(\emph{r})}$ and $\alpha _{a\emph{i}}^{(\emph{r})}$ by solving ${\bf{P}}_{4.23}$;}
\STATE {Update $\alpha _{b\emph{i}}^ {(\emph{r})} $ and $\alpha _{a\emph{i}}^ {(\emph{r})} $ ;}
\UNTIL{the objective function \eqref{31} converges in the ${\bf{P}}_{4.23}$;}
\STATE {Obtain $\alpha _{b\emph{i}}^ {*'} $ and $\alpha _{a\emph{i}}^ {*'} $};
\STATE{Update ${{\alpha _{b\emph{i}}^{(\gamma)}}}$, ${{\alpha _{a\emph{i}}^{(\gamma)}}}$};
\UNTIL{$\sum\limits_{\emph{i} = 1}^\emph{K} {\left( {R_s^{{\tau _\emph{i}}} + {\alpha _{b\emph{i}}}R_{s1}^{{t_\emph{i}}} + {\alpha _{a\emph{i}}}R_{s2}^{{t_\emph{i}}}} \right)}$ in \eqref{22} converges;}
\STATE{Obtain $\tau _\emph{i}^ * $, $t_\emph{i}^ * $, $\beta _\emph{i}^ * $, $P_{tr\emph{i}}^ * $, $\emph{T}_a^ *$ , $\emph{T}_b^ *$, $\alpha _{b\emph{i}}^ * $, $\alpha _{a\emph{i}}^ * $.}
\end{algorithmic}
\end{algorithm}

 The second sub-problem ${\bf{P}}_{4.2}$ contains integer variables $\alpha _{a\emph{i}}$ and $\alpha _{b\emph{i}}$, which can be optimally solved by the exhaustive search method (Brute-force), but this approach has high  computational complexity. To make the problem more tractable, we equivalently transform the integer constraint \eqref{22c} into continuous ones, given as
 \begin{align}
 {\rm{0}} \le {\alpha _{U\emph{i}}} \le 1,\forall \emph{i}{\rm{ }},\forall U \in \left\{ {a,b} \right\}, \label{26}
 \end{align}
 \begin{align}
 {\alpha _{U\emph{i}}} - \alpha _{U\emph{i}}^2 \le 0,\forall \emph{i},\forall U \in \left\{ {a,b} \right\}\label{27}.
 \end{align}
By replacing the constraint \eqref{22c} with \eqref{26} and \eqref{27}, we have
\begin{align}
{{\bf{P}}_{4.21}:}
&\mathop {\min }\limits_{{\alpha _{a\emph{i}}},{\alpha _{b\emph{i}}}} {\rm{ }}\sum\limits_{\emph{i} = 1}^K {-\left( { R{{_s^{{\tau _\emph{i}}}}^\prime } + {\alpha _{b\emph{i}}} R{{_{s1}^{{t_\emph{i}}}}^\prime } + {\alpha _{a\emph{i}}}R{{_{s2}^{{t_\emph{i}}}}^\prime }} \right)} \label{28}\\
{\rm{s}}.{\rm{t}}.
&{\rm{ }}\sum\limits_{\emph{i} = 1}^K {\left\{ {\hat R{{_p^{{\tau _\emph{i}}}}^\prime } + {\alpha _{b\emph{i}}}\hat R{{_{p1}^{{t_\emph{i}}}}^\prime } + {\alpha _{a\emph{i}}}R{{_{p2}^{{t_\emph{i}}}}^\prime }}\! -\! {R_0} \right\}} \ge \Delta ,\forall \emph{i}, \tag{28a}\label{28a}\\
&\eqref{22d}, \eqref{26}, \eqref{27}.  \tag{28b}\label{28b}
\end{align}

In ${{\bf{P}}_{4.21}}$, the objective function, constraints \eqref{22d}, \eqref{26} and \eqref{28a} are convex, yet, \eqref{27} is a reverse convex constraint due to the difference of two convex (d.c.) functions \cite{horst1999dc}, i.e., ${\alpha _{U\emph{i}}} - \alpha _{U\emph{i}}^2 $. Following \cite{6698281,6816086}, ${\bf{P}}_{4.21}$ can also be written as
\begin{align}\tag{29}\label{144}
 \mathop {\min }\limits_{\left( {{\alpha _{a\emph{i}}},{\alpha _{b\emph{i}}}} \right) \in \mathcal{D}} \mathop {\max }\limits_{\zeta  \ge 0} {\rm{ }}\;\;{\mathcal{L}}  \left( {{\alpha _{ai}},{\alpha _{bi}},\zeta } \right),
\end{align}
 where $\zeta$ is a non-negative constant, $\mathcal{D}$ is the feasible region satisfying all constraints, i.e., \eqref{22d}, \eqref{26}, \eqref{28a}, and $\mathcal{L}\left( {{\alpha _{a\emph{i}}},{\alpha _{b\emph{i}}},\zeta } \right)$ is given as
\begin{align}
\mathcal{L}\left( {{\alpha _{a\emph{i}}},{\alpha _{b\emph{i}}},\zeta } \right)=&{-\left( {R{{_s^{{\tau _\emph{i}}}}^\prime } + {\alpha _{b\emph{i}}}R{{_{s1}^{{t_\emph{i}}}}^\prime } + {\alpha _{a\emph{i}}}R{{_{s2}^{{t_\emph{i}}}}^\prime }} \right)}+\notag \\
&\zeta \sum\limits_{U \in \left\{ {a,b} \right\}} {\sum\limits_{\emph{i} = 1}^K {\left( {{\alpha _{U\emph{i}}} - \alpha _{U\emph{i}}^2} \right)} }.\tag{30}\label{145}
\end{align}
  The Lagrangian dual problem of \eqref{144} is expressed as
  \begin{align}\tag{31}\label{146}
  \mathop {\max }\limits_{\zeta  \ge 0}\mathop {\min }\limits_{\left( {{\alpha _{a\emph{i}}},{\alpha _{b\emph{i}}}} \right) \in \mathcal{D}} {\rm{ }}\;\;{\mathcal{L}}  \left( {{\alpha _{ai}},{\alpha _{bi}},\zeta } \right).
  \end{align}
   Generally speaking, there exists a nonzero duality gap between the primal problem and its duality \cite{boyd2004convex}, yet we can prove that the duality gap between \eqref{144} and its dual problem \eqref{146} equals zero, as summarized in {\bf{Lemma 2}}.

 {\bf{Lemma 2:}} The strong Lagrangian duality holds for \eqref{144} and its dual problem \eqref{146}, given as
 \begin{align}
 \mathop {\min }\limits_{\!\left( {{\alpha _{a\emph{i}}},{\alpha _{b\emph{i}}}} \!\right) \in \mathcal{D}} \mathop {\max }\limits_{\zeta  \ge 0} {\rm{ }}{\mathcal{L}} \! \left( {{\alpha _{ai}},{\alpha _{bi}},\zeta } \right)\! = \!\mathop {\max }\limits_{\zeta  \ge 0}\mathop {\min }\limits_{\!\left( {{\alpha _{a\emph{i}}},{\alpha _{b\emph{i}}}}\! \right) \in \mathcal{D}}\!\!\!\! {\rm{ }}{\mathcal{L}}  \!\left( \!{{\alpha _{ai}},{\alpha _{bi}},\zeta } \right).\tag{32}\label{D30}
 \end{align}
  \emph{Proof.} Please see Appendix D. \hfill {$\blacksquare $}

  Using {\bf{Lemma 2}}, we can transform ${\bf{P}}_{4.21}$ to
\begin{align}
{{\bf{P}}_{4.22}:}&\mathop {\min }\limits_{{\alpha _{a\emph{i}}},{\alpha _{b\emph{i}}}} {\rm{ }}\sum\limits_{\emph{i} = 1}^K {-\left( {R{{_s^{{\tau _\emph{i}}}}^\prime } + {\alpha _{b\emph{i}}}R{{_{s1}^{{t_\emph{i}}}}^\prime } + {\alpha _{a\emph{i}}}R{{_{s2}^{{t_\emph{i}}}}^\prime }} \right)}+\notag\\
&\;\;\;\;\;\;\;\;\;\;\;\zeta\!\! \sum\limits_{U \in \left\{ {a,b} \right\}} {\sum\limits_{\emph{i} = 1}^K {\left( {{\alpha _{U\emph{i}}} - \alpha _{U\emph{i}}^2} \right)} }\tag{33}\\
{\rm{s}}.{\rm{t}}. &\;\;\eqref{22d}, \eqref{26},\eqref{28a}. \tag{33a}
\end{align}
\!Please note that for a sufficiently large  $\zeta$, the variables ${\alpha _{a\emph{i}}}$ and ${\alpha _{b\emph{i}}}$ will be forced to approach 0 or 1, thereby satisfying the constraint (27) in ${\bf{P}}_{4.21}$.

 ${{\bf{P}}_{4.22}}$ is still non-convex due to the non-convex term $\sum\limits_{U \in \left\{ {a,b} \right\}} {\sum\limits_{\emph{i} = 1}^K {\left( {\alpha _{U\emph{i}} - \alpha _{U\emph{i}}^2} \right)} } $ in the objective function. In order to handle it, the first-order Taylor expansion is used to find its upper bound, which is given as
\begin{align}
&\sum\limits_{U \in \left\{ {a,b} \right\}} {\sum\limits_{\emph{i} = 1}^K {\left( {{\alpha _{U\emph{i}}} - \alpha _{U\emph{i}}^2} \right)} }  \le \sum\limits_{U \in \left\{ {a,b} \right\}}  \sum\limits_{\emph{i} = 1}^K  \left[ {{\alpha _{U\emph{i}}} - {{\left( {\alpha _{U\emph{i}}^{(\emph{r})}} \right)}^2}} \right. \notag \\
&\;\;\;\;\;\;\;\;\;\;\;\;\;\;\left. { - 2\alpha _{U\emph{i}}^{(\emph{r})}\left( {{\alpha _{U\emph{i}}} - \alpha _{U\emph{i}}^{(\emph{r})}} \right)} \right] \buildrel \Delta \over =  M\left( \alpha  \right), \tag{34} \label{29}
\end{align}
where $\alpha _{U\emph{i}}^{(\emph{r})}$ denotes the obtained value after \emph{r}-th iteration.

Using $M\left( \alpha  \right)$ to replace the non-convex term $\sum\limits_{U \in \left\{ {a,b} \right\}} {\sum\limits_{\emph{i} = 1}^K {\left( { \alpha _{U\emph{i}}- \alpha _{U\emph{i}}^2} \right)} } $, ${{\bf{P}}_{4.22}}$ can be transformed into the convex one, i.e.,
\begin{align}
{{\bf{P}}_{4.23}:}&\mathop {\min }\limits_{{\alpha _{a\emph{i}}},{\alpha _{b\emph{i}}}} {\rm{ }}\!\!\sum\limits_{\emph{i} = 1}^K {\!-\!\left(\! {R{{_s^{{\tau _\emph{i}}}}^\prime }\! +\! {\alpha _{b\emph{i}}}R{{_{s1}^{{t_\emph{i}}}}^\prime }\! +\! {\alpha _{a\emph{i}}}R{{_{s2}^{{t_\emph{i}}}}^\prime }} \right)}\!+\! \zeta M\left(\! {\alpha}\! \right)   \tag{35}\label{31}\\
{\rm{s}}.{\rm{t}}. \;\;\;&\eqref{22d}, \eqref{26}, \eqref{28a}.  \tag{35a}\label{31a}
\end{align}

Based on the above, the original optimization problem is decomposed into two subproblems, each of which is subsequently converted into a convex form and a BCD-based iterative algorithm is developed to solve the transformed problems.

\section{Complexity and Convergence Analysis}
\subsection{Complexity Analysis of the Proposed Algorithms}
In this subsection, we analyze the complexity of Algorithm 1 and Algorithm 2. Assume that the interior point method is used, the computational complexity of Algorithm \ 1 for solving ${{\bf{P}}_3} $ is $O\left( \emph{I}_1{\sqrt {\emph{M}_1} \frac{1}{{\delta}_1}} \right)$ \cite{boyd2004convex,gondzio1994computational}, where ${\delta}_1$ is iterative accuracy, $\emph{I}_1$ signifies the iterations needed for the convergence and $\emph{M}_1=6\emph{K}$ denotes the number of optimization variables, which is related to the number of SUs. For BCD-based algorithm,
the computational complexity for solving sub-problem ${\bf{P}}_{4.11}$ and ${\bf{P}}_{4.23}$ is $O\left( \emph{I}_m{\sqrt {\emph{M}_m} \frac{1}{{\delta}_m}} \right)$, where $m \in \left\{ {2, 3} \right\}$ corresponds to
${\bf{P}}_{4.11}$ and ${\bf{P}}_{4.23}$, respectively, $\emph{I}_m$ is the number of iterations required for convergence, ${\delta}_m$ is the iterative accuracy and $\emph{M}_m$ denotes the number of optimization variables. Specifically, $\emph{M}_2=6\emph{K}$, $\emph{M}_3=2\emph{K}$.
 Thus, Algorithm \ 2 has a computational complexity of $O\left[ {{\emph{I}_4}\left( {{\emph{I}_2}\sqrt {{\emph{M}_2}} \frac{1}{{{\delta _2}}} + {\emph{I}_3}\sqrt {{\emph{M}_3}} \frac{1}{{{\delta _3}}}} \right)} \right]$, where ${\emph{I}_4}$ is the number of iterations required for convergence of solving  (22). In practice IoT implementations, the number of IoT nodes (i.e., SUs) is inherently limited to avoid the intense resource competition. Therefore, in practical scenarios, the proposed algorithm can effectively maximize SUs' total rate while ensuring low computational complexity.

\subsection{ Convergence Analysis for Algorithms 1 and 2}
The core idea of the SCA method is to replace non-convex terms with their upper or lower bounds, thereby transforming the original problem into a more tractable form and obtaining suboptimal solutions. The convergence of the SCA method has been theoretically proved in \cite{razaviyayn2014successive}.
In our work, Algorithm 1 is developed based on SCA. Specifically, we approximate the non-convex terms $f(P_{tr{\emph{i}}})$ in the ${\bf{P}}_1$ using its lower bound. Through further transformation, the optimization problem ${\bf{P}}_3$ is formulated. Therefore, the feasible set of ${\bf{P}}_3$ becomes a subset of that in ${\bf{P}}_1$, and ${\bf{P}}_3$'s objective function serves as a lower bound for ${\bf{P}}_1$.
Based on the convergence of SCA, the convergence of Algorithm 1 is theoretically guaranteed.

Algorithm 2 is proposed to solving ${\bf{P}}_4$ by means of the BCD method, its convergence analysis is proved as follows. Firstly, define $R_{total}$ as the total rate of SUs in ${\bf{P}}_{4}$, and $R_{total,lb}$ as its lower bound  due to the first-order Taylor approximation of the non-convex term. Let $\left\{ {\tau _{\emph{i}}^{\left( u \right)},t_{\emph{i}}^{\left( u \right)},\beta _{\emph{i}}^{\left( u \right)},P_{tr{\emph{i}}}^{\left( u \right)},\emph{T}_a^{\left( u \right)},\emph{T}_b^{\left( u \right)},\alpha _{b{\emph{i}}},\alpha _{a{\emph{i}}}} \right\}$ denote the solution at the $u$-th iteration for solving ${\bf{P}}_{4.11}$, and $\left\{ {\tau _{\emph{i}}^{\left( {u + 1} \right)},t_{\emph{i}}^{\left( {u + 1} \right)},\beta _{\emph{i}}^{\left( {u + 1} \right)},P_{tr{\emph{i}}}^{\left( {u + 1} \right)},\emph{T}_a^{\left( {u + 1} \right)}},\emph{T}_b^{\left( {u + 1} \right)}, \right.$ $\left. {\alpha _{b{\emph{i}}}^{\left( {r + 1} \right)},\alpha _{a{\emph{i}}}^{\left( {r + 1} \right)}} \right\}$ is the solution obtained in the $\left( {r + 1} \right)$-th iteration for solving ${\bf{P}}_{4.23}$. Then, we have
\begin{align}\label{36}\tag{36}
\begin{array}{l}
{R_{total}}\left( {\tau _{\emph{i}}^{\left( u \right)},t_{\emph{i}}^{\left( u \right)},\beta _{\emph{i}}^{\left( u \right)},P_{tr{\emph{i}}}^{\left( u \right)},T_a^{\left( u \right)},T_b^{\left( u \right)},{\alpha _{a{\emph{i}}}},{\alpha _{b{\emph{i}}}}} \right)\\
\mathop  = \limits^{\left( a \right)} {R_{total,lb}}\left( {\tau _{\emph{i}}^{\left( u \right)},t_{\emph{i}}^{\left( u \right)},\beta _{\emph{i}}^{\left( u \right)},P_{tr{\emph{i}}}^{\left( u \right)},T_a^{\left( u \right)},T_b^{\left( u \right)},{\alpha _{a{\emph{i}}}},{\alpha _{b{\emph{i}}}}} \right)\\
\mathop  \le \limits^{\left( b \right)}\!\! {R_{total,lb}}\!\!\left(\! {\tau _{\emph{i}}^{\!\left(\! {u + 1}\! \right)}\!,t_{\emph{i}}^{\left( \!{u + 1}\! \right)}\!,\beta _{\emph{i}}^{\!\left( \!{u + 1}\! \right)}\!,P_{tr{\emph{i}}}^{\!\left(\! {u + 1} \!\right)}\!,\emph{T}_a^{\left( \!{u + 1}\! \right)}\!,\emph{T}_b^{\left( \!{u + 1}\! \right)}\!,\!{\alpha _{a{\emph{i}}}},\!{\alpha _{b{\emph{i}}}}}\! \right)\\
\mathop  \le \limits^{\left( c \right)}\!\! {R_{total,lb}}\!\!\left(\!\! {\tau _{\emph{i}}^{\!\left( \!{u + 1} \! \right)}\!,\!t_{\emph{i}}^{\!\left( \!{u + 1}\! \right)}\!,\!\beta _{\emph{i}}^{\!\left(\! {u + 1}\! \right)}\!,\!P_{tr{\emph{i}}}^{\!\left(\! {u + 1}\! \right)}\!,\!\emph{T}_a^{\left( {u + 1}\! \right)}\!,\!\emph{T}_b^{\!\left(\! {u + 1}\! \right)}\!,\!\alpha _{b{\emph{i}}}^{\!\left(\! {r + 1}\! \right)}\!,\!\alpha _{a{\emph{i}}}^{\!\left( \!{r + 1}\! \right)}} \!\!\right)\\
\mathop  \le \limits^{\left( d \right)}\!\! {R_{total}}\!\!\left(\!\! {\tau _{\emph{i}}^{\!\left( \!{u + 1} \! \right)},\!t_{\emph{i}}^{\!\left( \!{u + 1}\! \right)},\!\beta _{\emph{i}}^{\!\left(\! {u + 1}\! \right)},\!P_{tr{\emph{i}}}^{\!\left(\! {u + 1}\! \right)},\!\emph{T}_a^{\!\left( \!{u + 1}\! \right)},\!\emph{T}_b^{\!\left(\! {u + 1}\! \right)},\!\alpha _{b{\emph{i}}}^{\!\left(\! {r + 1}\! \right)},\!\alpha _{a{\emph{i}}}^{\!\left( \!{r + 1}\! \right)}} \!\!\right)
\end{array},
\end{align}
where $\left( a \right)$ comes from the fact that the Taylor expansion in
 $f(P_{tr{\emph{i}}})$ is tight at the given local point $P_{tr{\emph{i}}}^{\left( u \right)}$; $\left( b \right)$ holds because ${\bf{P}}_{4.11}$ has a locally optimal solution at the new point; $\left( c \right)$ holds since $\left\{ {\alpha _{b{\emph{i}}}^{\left( {r + 1} \right)},\alpha _{a{\emph{i}}}^{\left( {r + 1} \right)}} \right\}$ is locally optimal for solving ${\bf{P}}_{4.23}$; $\left( d \right)$ always holds due to the fact that the maximal SUs' total rate of ${\bf{P}}_{4}$ is lower bounded by that of ${\bf{P}}_{4.23}$ \cite{9222571}. The above inequality implies that the SUs' total rate is always non-decreasing with each iteration. Moreover, the objective value of problem ${\bf{P}}_{4}$ is upper bounded due to the boundedness of the optimization variables. Hence, the convergence of Algorithm 2 can be theoretically ensured.

 \section{Numerical Results}
%
In this section, simulation results are presented to demonstrate the performance of the proposed mutualistic SR with HAPC over the traditional one. Unless otherwise stated, the basic parameters are set as follows.
According to \cite{9686018,10463656,9534878},
the locations of the PT, SU${_1}$, SU${_2}$ and the receiver are set to (0,0), (0.8 m,0), ($-$0.8 m,0) and (0,100 m), respectively, the channel bandwidth $W$=10 KHz, the noise power spectral density is $-$90 dBm/Hz, the energy conversion efficiency is 0.8, the backscatter circuit power consumption and active circuit power consumption are ${10^{ - 5}}$w and ${10^{ - 3}}$w separately. Besides, the channel gains are set as ${\xi _{\emph{i}}}h_{\emph{i}}^{ - {\alpha _{\emph{i}}}}$,
where ${\xi _\emph{i}}$, ${\alpha _\emph{i}}$ and ${h_\emph{i}}$ denote the power gain of the small-scale fading, the path loss exponent and the distance for the \emph{i}-th link, $\emph{i} \in \left\{ {{\rm PR},{\rm PS_1},{\rm S_1R},{\rm PS_{\rm{S_1R}}},{\rm S_2R},{\rm S_{12}}} \right\}$. Moreover, we have ${\alpha _{\rm{PR}}} = 3.5$, ${\alpha _{\rm{{PS}_1}}} = 2.5$, ${\alpha _{\rm{S_1R}}} = 2.9$, ${\alpha _{\rm{PS_2}}} = 2.5$, ${\alpha _{\rm{S_2R}}} = 2.9$, ${\alpha _{\rm{S_{12}}}} = 2.5$.
For a fair comparison, we define the baseline as the traditional mutualistic SR, where each SU performs only passive BC in TDMA mode. Specifically, each SU sequentially performs BC and harvests energy from the PT's signal during its inactive period. This process aligns with the BC phase as previously described in Section II-A.
%
  \begin{figure*}[t]
  \begin{minipage}[t]{0.495\linewidth}
\centering
\includegraphics[width=0.86\textwidth]{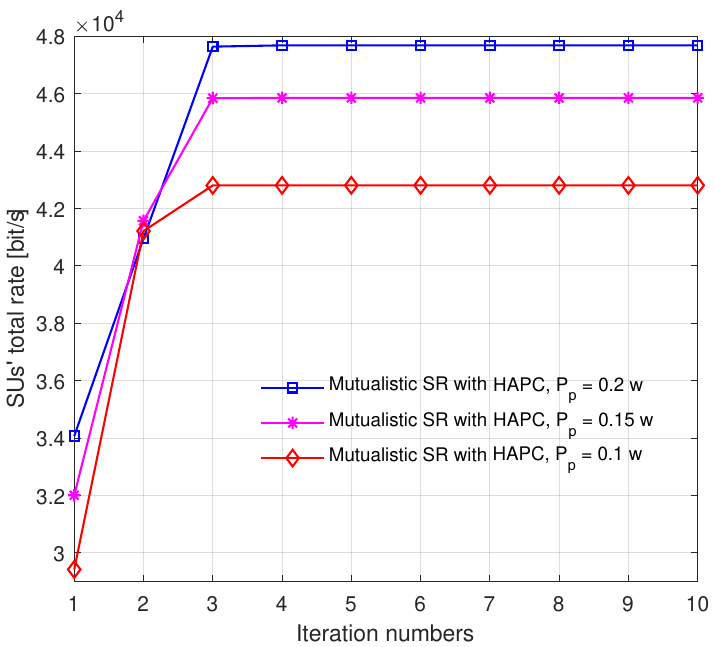}
\centering \caption{SUs' total rate versus iteration numbers.}
\end{minipage}
\begin{minipage}[t]{0.495\linewidth}
\centering
\includegraphics[width=0.86\textwidth]{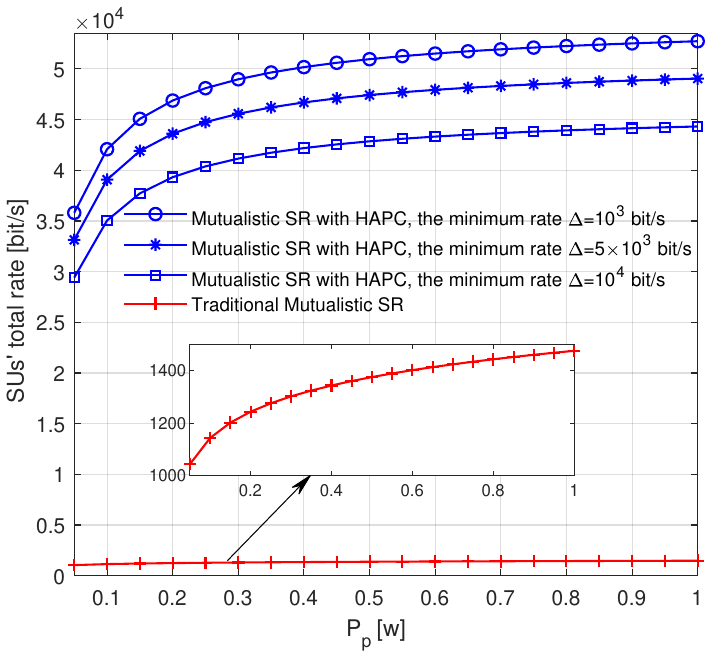}
\centering \caption{SUs' total rate versus ${P_p}$.}
\end{minipage}
\end{figure*}
  \begin{figure*}[t]
  \begin{minipage}[t]{0.495\linewidth}
\centering
\includegraphics[width=0.86\textwidth]{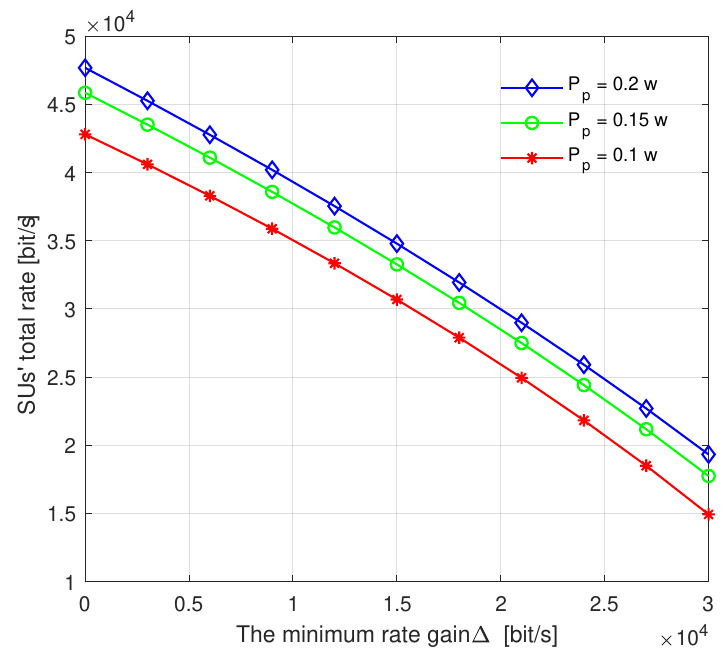}
\centering \caption{SUs' total rate versus  $\Delta $.}
\end{minipage}
\begin{minipage}[t]{0.495\linewidth}
\centering
\includegraphics[width=0.86\textwidth]{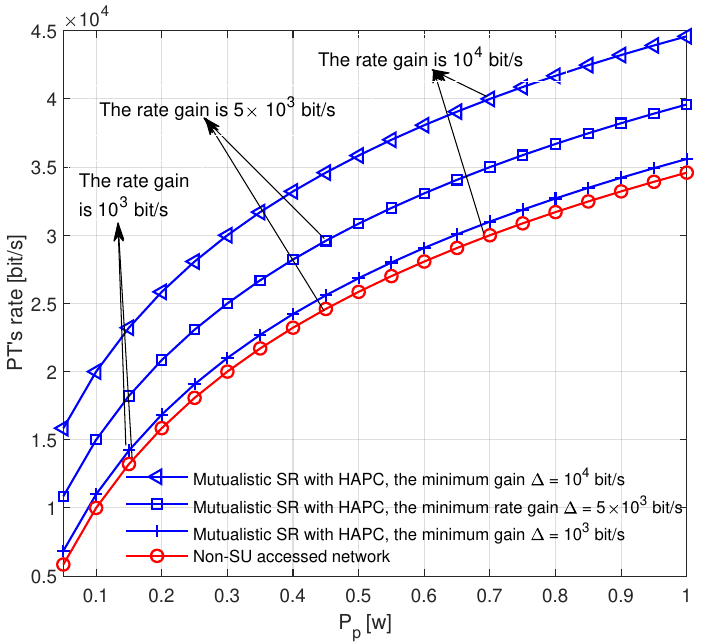}
\centering \caption{PT's rate versus $P_p $.}
\end{minipage}
\end{figure*}

The convergence of Algorithm 1 is shown in Fig. 2. It can be noticed that the algorithm quickly converges after four or five iterations. With the change of PT's transmit power, the convergence of the algorithm does not change, which indicates that the proposed algorithm has a stable convergence performance. Besides, the total rate of SUs increases with the increase of PT's transmit power. The reason is that a higher PT's transmit power allows each SU to harvest more energy for AC and thus achieves a higher rate.

  \begin{figure*}[t]
  \begin{minipage}[t]{0.495\linewidth}
\centering
\includegraphics[width=0.86\textwidth]{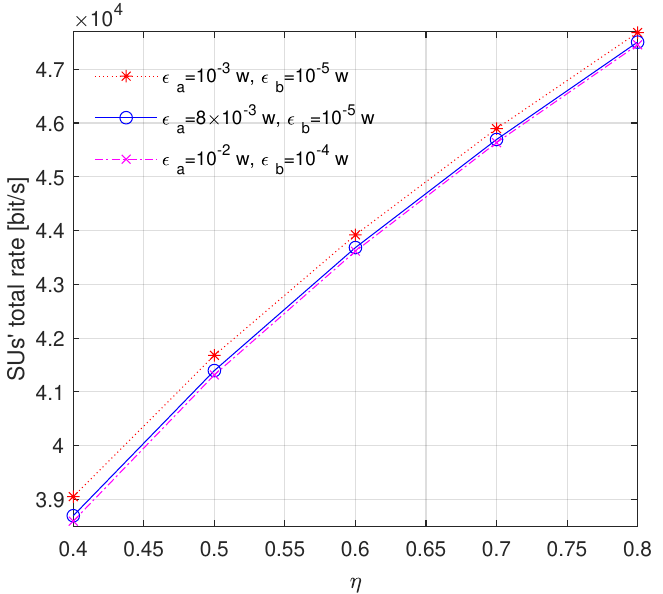}
\centering \caption{SUs' total rate vs $\eta$.}
\end{minipage}
\begin{minipage}[t]{0.495\linewidth}
\centering
\includegraphics[width=0.86\textwidth]{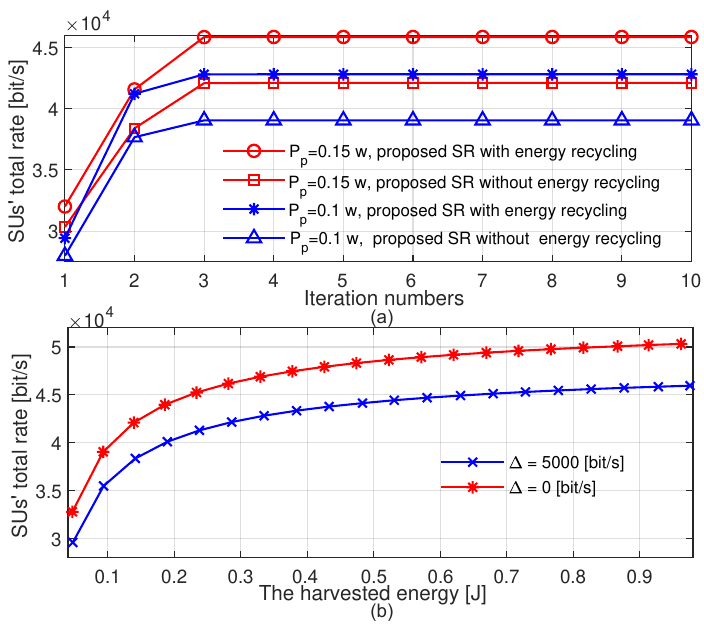}
\centering \caption{SUs' total rate versus iteration numbers and the harvested energy.}
\end{minipage}
\end{figure*}

   \begin{figure*}[t]
  \begin{minipage}[t]{0.495\linewidth}
\centering
\includegraphics[width=0.86\textwidth]{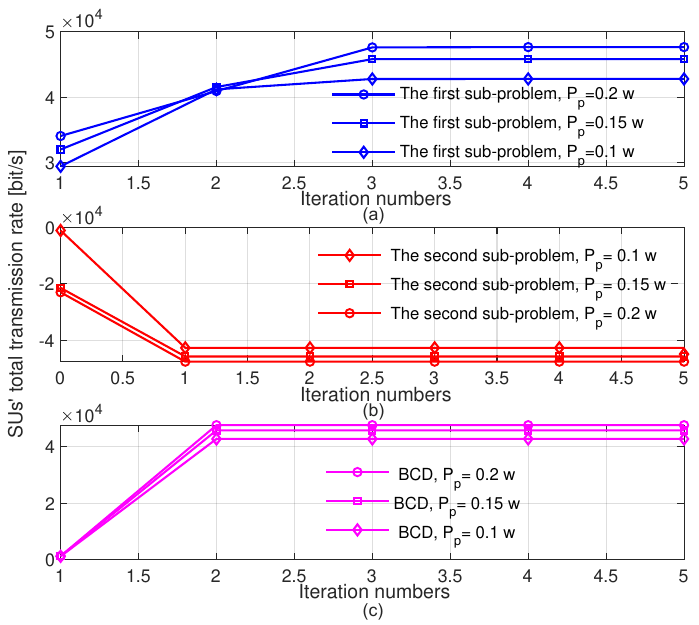}
\centering \caption{Convergence illustration of the algorithm 2.}
\end{minipage}
\begin{minipage}[t]{0.495\linewidth}
\centering
\includegraphics[width=0.86\textwidth]{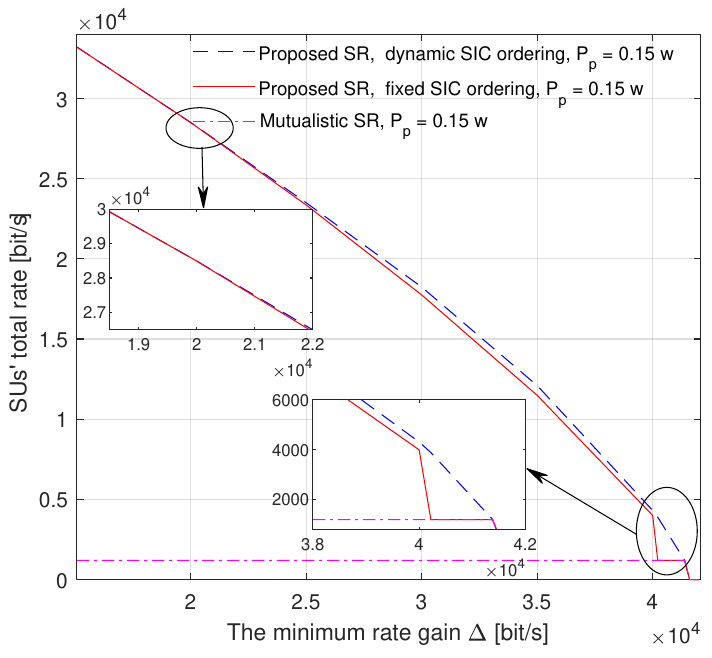}
\centering \caption{The comparison of SUs' total rate with fixed/dynamic SIC ordering.}
\end{minipage}
\end{figure*}

%

Fig.3 compares the total rate of SUs achieved in our proposed scheme with that achieved in traditional mutualistic SR. It is evident that our proposed network achieves a better performance than the traditional setup in terms of SUs' total rate. Specifically, SUs' total rate in our proposed scheme is nearly 40 times higher than that of the traditional SR when $\Delta=10^3$ bit/s, which indicates that HAPC does benefit in improving SUs' rate. Besides, we note that with the increase of PT's minimum rate gain, a consequent decrease in SUs' total rate occurs. This is because the increase in PT's minimum rate gain is accompanied by a decrease in communication resources for the SUs to perform AC, resulting in a relatively lower SUs' total rate. Nevertheless, the rate of our proposed network is still greater than that of the traditional mutualistic SR.

To further analyze the impacts of PT's minimum rate gain (i.e., $\Delta$) on the total rate of SUs, Fig. 4 shows the variation curves of SUs' total rate versus $\Delta$ for different $P_p$. We can see that the SUs' total rate shows a decreasing trend with the increasing of PT's minimum rate gain under different PT's transmit power. The reason is summarized below. As $\Delta$ increases, the system must allocate more resources to the BC phase to satisfy the constraint ${\Delta _1} + {\Delta _2} \ge \Delta$ in ${\bf{P}}_1$, ensuring sufficient rate gain for the PT. However, this comes at the cost of reduced resources for the AC phase. Since the rate achieved during the AC phase plays a dominant role in determining the SUs' total rate, this shift leads to a decline in the overall transmission performance of the SUs.


Fig. 5 depicts the variation of PT's rate with $P_p$ for different schemes. The rate of PT shows an increasing trend with the increase of $P_p$. Specifically, when PT's minimum rate gain is $10^3$ bit/s, the rate gain, which is defined as  the difference between the  PT's rate under our proposed network and  the  network where all  SUs are not admitted to access PT's spectrum resource, is equal to PT's minimum rate gain. When PT's minimum rate increases by $5\times10^3$ bit/s, the corresponding rate gain reaches  $5\times10^3$ bit/s. The same situation occurs when PT's minimum rate gain is $10^4$ bit/s. The above observations indicate that in our simulation parameters, the equality in constraint \eqref{13a} holds for maximizing the total rate of SUs, and also agree with Proposition 1.

As shown in Fig. 6, SUs' total rate increases with higher values of $\eta$. This is because a larger $\eta$ allows the SUs to harvest more energy, enabling them to transmit with higher power during active communication (AC) phase, which in turn improves the overall total rate. Under the same $\eta$, a high circuit power consumption leads to a reduction in the rate of the SUs. This is due to the fact that more harvested energy is consumed to support circuit operation, reducing the energy available for data transmission and thereby limiting the transmission rate.

Fig. 7(a) shows the advantages of the energy recycling. We can clearly observe that the SUs' total rate in the case where energy recycling is considered is always higher than the scenario where energy recycling is not considered. In Fig. 7(b), we solve the problem at different PT's transmit power. For a given PT's transmit power, we depict the harvested energy on the horizontal axis and the total rate of SUs on the vertical axis in Fig. 7(b), respectively. It can be observed that the total rate of SUs increases with the harvested energy across different minimum rate gains of PT.
 This is due to the fact that the more energy SUs collect, the more opportunities SUs have for longer transmission time or higher transmit power during AC phase, thereby leading to a higher rate. Hence, considering energy recycling is an efficient way to enhance efficiency of wireless power transfer, especially in scenarios with dense SU deployments.
%

Fig. 8 shows the convergence of Algorithm 2, where Fig. 8(a), Fig. 8(b) and Fig. 8(c) portray the convergence of ${\bf{P}}_{4.1}$, ${\bf{P}}_{4.2}$ and ${\bf{P}}_{4}$, respectively. We can see that the SUs' total rate in Fig. 8(a)$-$(c) converge quickly after three iterations under different maximum PT's transmit power, which indicates that our proposed Algorithm has strong convergence properties.

Fig. 9 demonstrates performance difference of the total rate with fixed/dynamic SIC ordering. Firstly, it can be observed that the proposed scheme achieves a higher rate compared to the traditional mutualistic SR, regardless of fixed SIC ordering or dynamic SIC ordering. In addition, when $\Delta$ is small, the fixed SIC ordering can achieve the same performance as the dynamic SIC ordering. When $\Delta$  continues to increase, the dynamic SIC ordering can achieve a better performance. It can be explained below. For a large $\Delta$, if the fixed SIC ordering is considered, to ensure PT's  minimum rate gain, the SU has to always operate in the BC mode and no time is allocated for AC. While for the dynamic ordering, decoding SU$_\emph{i}$'s signal leads to a zero PT's rate loss and this may let  SU$_\emph{i}$'s operate in both BC and AC modes.

\section{Conclusion}
This paper has introduced a novel mutualistic SR with HAPC, in which each SU transmits information via BC and AC alternately. We have formulated two optimization problems with the objective of maximizing the total rate of all SUs under the fixed SIC ordering and the dynamic SIC ordering, respectively, and have also developed two iterative algorithms to solve them. For the fixed SIC ordering, the problem was solved by jointly using SCA and auxiliary variables. For the dynamic SIC ordering, we have transformed the MIP problem into a convex one by means of a series of techniques such as BCD, SCA and auxiliary variables.
Simulation results have confirmed the strong convergence capabilities of the proposed algorithms and the following facts. First, compared to the traditional mutualistic SR, the proposed network can achieve a significantly higher total rate for SUs. Second, SUs' total rate under the dynamic SIC ordering is larger than that of the fixed one when the PT's minimum rate gain is high, and becomes nearly identical when the PT's minimum rate gain is low. Our work can be extended to more practical scenarios in the future. Specifically, robust optimization or stochastic programming can be employed to handle imperfect CSI. To adapt to dynamic environments such as user mobility, adaptive algorithms may be developed. Additionally, decentralized optimization architectures can be explored to enhance scalability and robustness in dense IoT networks.

\appendices
\begin{appendices}
\section{ }
  Here we prove that $W{\tau _\emph{i}}{{\mathbb{E}}_{{c_\emph{i}}}}\left[ {{{\log }_2}\left( {1 + \frac{{{P_p}{{\left| {\sqrt b  + \sqrt {{{{a_\emph{i}}{d_\emph{i}}{q_\emph{i}}} \mathord{\left/ {\vphantom {{{a_\emph{i}}{d_\emph{i}}{q_\emph{i}}} {{\tau _\emph{i}}}}} \right. \kern-\nulldelimiterspace} {{\tau _\emph{i}}}}} {c_\emph{i}}} \right|}^2}}}{{{\sigma ^2}}}} \right)} \right]$ is a jointly concave function with respect to ${q_\emph{i}}$ and ${\tau _\emph{i}}$. Denote $W{\tau _\emph{i}}{{\mathbb{E}}_{{c_\emph{i}}}}\left[ {{{\log }_2}\left( {1 + \frac{{{P_p}{{\left| {\sqrt b  + \sqrt {{{{a_\emph{i}}{d_\emph{i}}{q_\emph{i}}} \mathord{\left/ {\vphantom {{{a_\emph{i}}{d_\emph{i}}{q_\emph{i}}} {{\tau _\emph{i}}}}} \right. \kern-\nulldelimiterspace} {{\tau _\emph{i}}}}} {c_\emph{i}}} \right|}^2}}}{{{\sigma ^2}}}} \right)} \right]$ as ${\nabla _1}\!\left( {{q_\emph{i}}, {\tau _\emph{i}}} \right)$. Since the expectation function preserves convexity \cite{boyd2004convex}, the convexity of ${\nabla _1}\!\left( {{q_\emph{i}}, {\tau _\emph{i}}} \right)$ is the same as that of ${\nabla _2}\!\left( {{q_\emph{i}},\!{\tau _\emph{i}}} \right)\!\! =\!\!W\!{\tau _\emph{i}}{\log _2}\!\!\left( \!\!{1\!\! +\! \frac{{{P_p}\!{{\left| \!{\sqrt b\!  +\! \sqrt {{{{\!a_\emph{i}}{d_\emph{i}}{q_\emph{i}}} \mathord{\left/ {\vphantom {{\!\!{a_\emph{i}}{d_\emph{i}}{q_\emph{i}}} {{\tau _\emph{i}}}}}\! \right.
 \kern-\nulldelimiterspace} {{\tau _\emph{i}}}}} {c_\emph{i}}} \right|}^2}}}{{{\sigma ^2}}}} \!\!\right)$.  As ${\nabla _2}\left( {{q_\emph{i}}, {\tau _\emph{i}}} \right)$ is a perspective function of ${\nabla _3}\left( {{q_\emph{i}}} \right) = W{\log _2}\left( {1 + \frac{{{P_p}{{\left| {\sqrt b  + \sqrt {{a_\emph{i}}{d_\emph{i}}{q_\emph{i}}} {c_\emph{i}}} \right|}^2}}}{{{\sigma ^2}}}} \right)$, ${\nabla _2}$ and  ${\nabla _3}$ have the same  convexity \cite{boyd2004convex}, and we only need to justify the convexity of ${\nabla _3}$. Clearly,  ${\nabla _3}$ is concave with respect to $q_\emph{i}$, which indicates   ${\nabla _1}\left( {{q_\emph{i}}, {\tau _\emph{i}}} \right)$ is concave and the proof is complete.

\section{ }
Proving that at least one of the equations in (13a) and (13b) holds is equivalent to proving that ${\Delta _1} + {\Delta _2} > \Delta$ and $H_{b\emph{i}}^{har} + H_{a\emph{i}}^{har} > H_{b\emph{i}}^{con} + H_{a\emph{i}}^{con}, \forall i$ cannot simultaneously hold. Keeping this in mind, the proof by contradiction is adopted. Assume that the optimal solution to the optimization problem ${{\bf{P}}_1}$ is $\left( {\tau _\emph{i}^*,t_\emph{i}^*,\beta _\emph{i}^*,P_{tr\emph{i}}^*,\emph{T}_a^ * ,\emph{T}_b^ * } \right)$, which satisfies all constraints and the following inequalities, ${\Delta _1} + {\Delta _2} > \Delta $ and $H_{b\emph{i}}^{har} + H_{a\emph{i}}^{har} > H_{b\emph{i}}^{con} + H_{a\emph{i}}^{con}, \forall \emph{i}{\rm{ }}$. The corresponding objective function value is $R\left( {\tau _\emph{i}^* ,t_\emph{i}^* ,\beta _\emph{i}^* ,P_{tr\emph{i}}^* ,\emph{T}_a^* ,\emph{T}_b^* } \right)$.

We construct another feasible solution $\left( {\tau _\emph{i}^\dag ,t_\emph{i}^\dag ,\beta _\emph{i}^\dag ,P_{tr\emph{i}}^\dag ,\emph{T}_a^\dag ,\emph{T}_b^\dag } \right)$ that satisfies all constraints and ${\Delta _1} + {\Delta _2} > \Delta $ and $H_{b\emph{i}}^{har} + H_{a\emph{i}}^{har} = H_{b\emph{i}}^{con} + H_{a\emph{i}}^{con}$. The corresponding solution is denoted as $R\left( {\tau _\emph{i}^\dag ,t_\emph{i}^\dag ,\beta _\emph{i}^\dag ,P_{tr\emph{i}}^\dag ,\emph{T}_a^\dag ,\emph{T}_b^\dag } \right)$.  At the same time, the above two sets of solutions satisfy $\tau _\emph{i}^* = \tau _\emph{i}^\dag $, $t_\emph{i}^* = t_\emph{i}^\dag $, $P_{tr\emph{i}}^* = P_{tr\emph{i}}^\dag $, $\emph{T}_a^ *  = \emph{T}_a^\dag $, $\emph{T}_b^ *  = \emph{T}_b^\dag $, and $\beta _\emph{i}^* < \beta _\emph{i}^\dag  < 1$.

Clearly, the objective function \eqref{13} is an increasing function about ${\beta _\emph{i}}$. Thus, the following relationship
$R\!\!\left( \!{\tau _\emph{i}^\dag ,\!t_\emph{i}^\dag ,\!\beta _\emph{i}^\dag ,\!P_{tr\emph{i}}^\dag ,\!\emph{T}_a^\dag ,\!\emph{T}_b^\dag }\! \right)> \!R\left( \!{\tau _\emph{i}^*,\!t_\emph{i}^*,\!\beta _\emph{i}^*,\!P_{tr\emph{i}}^*,\!\emph{T}_a^ * ,\!\emph{T}_b^ * } \right)$ exists, which obviously contradicts to the original hypothesis. Thus, inequalities in (13a) and (13b) cannot hold simultaneously. That is, at least one equality holds in constraints (13a) and (13b).

 \section{ }
 By means of contradiction, we prove that the equality of constraint (13h) must hold. Assume that the optimal solution of the ${{\bf{P}}_{1}}$ is $\left( {\tau _{\emph{i}}^ * ,t_{\emph{i}}^ * ,\beta _{\emph{i}}^ * ,P_{tr{\emph{i}}}^ * ,\emph{T}_a^ * ,\emph{T}_b^ * } \right)$, which satisfies all constraints and $\emph{T}_a^ * + \emph{T}_b^ *  <  \emph{T}$, i.e., the allocated time is not completely used up. The corresponding objective function is denoted as $R\left( {\tau _{\emph{i}}^ * ,t_{\emph{i}}^ * ,\beta _{\emph{i}}^ * ,P_{tr{\emph{i}}}^ * ,\emph{T}_a^ * ,\emph{T}_b^ * } \right)$.

 Then, we construct another set of feasible solutions $\left( {\tau _{\emph{i}}^\dag ,t_{\emph{i}}^\dag ,\beta _{\emph{i}}^\dag ,P_{tr{\emph{i}}}^\dag  ,\emph{T}_a^\dag ,\emph{T}_b^\dag } \right)$, which satisfies all constraints and $\emph{T}_a^\dag + \emph{T}_b^\dag = \emph{T}$, i.e., the allocated energy is used up. At the same time, we have $P_{tr{\emph{i}}}^ *  = P_{tr{\emph{i}}}^\dag$, $\beta _{\emph{i}}^ *  = \beta _{\emph{i}}^\dag $, $\emph{T}_a^ *  <\emph{T}_a^\dag $, $\emph{T}_b^ * < \emph{T}_b^\dag $. Since the total transmission time allocated to all SUs for AC and BC are respectively increased, there must exist at least one set of $t_{\emph{i}}^\dag$ and $\tau _{\emph{i}}^\dag$ satisfying $t_{\emph{i}}^ *  < t_{\emph{i}}^\dag $, $\tau _{\emph{i}}^ *  < \tau _{\emph{i}}^\dag $. The corresponding objective function is denoted as $R\left( {\tau _{\emph{i}}^\dag ,t_{\emph{i}}^\dag ,\beta _{\emph{i}}^\dag ,P_{tr{\emph{i}}}^\dag,\emph{T}_a^\dag ,\emph{T}_b^\dag } \right)$. Since the objective function is an increasing function with respect to $t_{\emph{i}}$ and $\tau _{\emph{i}}$. Therefore, we have $R\left( {\tau _{\emph{i}}^ * ,t_{\emph{i}}^ * ,\beta _{\emph{i}}^ * ,P_{tr{\emph{i}}}^ * ,\emph{T}_a^ * ,\emph{T}_b^ * } \right) < R\left( {\tau _{\emph{i}}^\dag ,t_{\emph{i}}^\dag ,\beta _{\emph{i}}^\dag ,P_{tr{\emph{i}}}^\dag ,T_a^\dag ,\emph{T}_b^\dag } \right)$, which is not consistent with the original hypothesis. Thus, there must exist  $\emph{T}_a^\dag + \emph{T}_b^\dag = \emph{T}$ such that all of the allocated time is used up and the objective function is maximized.

\section{ }

According to the Lagrange duality theorem, we have the following inequality
\begin{align}
{d^ * }=&\mathop {\min }\limits_{\left( {{\alpha _{a\emph{i}}},{\alpha _{b\emph{i}}}} \right) \in \mathcal{D}} \mathop {\max }\limits_{\zeta  \ge 0} {\rm{ }}\;\;{\mathcal{L}}  \left( {{\alpha _{ai}},{\alpha _{bi}},\zeta } \right) \ge \notag \\
 &\mathop {\max }\limits_{\zeta  \ge 0}\mathop {\min }\limits_{\left( {{\alpha _{a\emph{i}}},{\alpha _{b\emph{i}}}} \right) \in \mathcal{D}} {\rm{ }}\;\;{\mathcal{L}}  \left( {{\alpha _{ai}},{\alpha _{bi}},\zeta } \right)=\mathop {\max }\limits_{\zeta  \ge 0}\mathop {\varpi \left( \zeta  \right)},\tag{D.1}\label{D1}
\end{align}
where ${d^ * }$ is the optimal value of ${\bf{P}}_{4.21}$, $\varpi \left( \zeta  \right) \buildrel \Delta \over = \mathop {\min }\limits_{\left( {{\alpha _{a\emph{i}}},{\alpha _{b\emph{i}}}} \right) \in D} L\left( {{\alpha _{a\emph{i}}},{\alpha _{b\emph{i}}},\zeta } \right)$. In the following, we can prove that strong duality still holds. When $\left( {{\alpha _{a\emph{i}}},{\alpha _{b\emph{i}}}} \right) \in \mathcal{D}$, ${\alpha _{U\emph{i}}} - \alpha _{U\emph{i}}^2 \ge 0$ is met, ${\varpi \left( \zeta  \right)}$ is a monotonically increasing function with respect to $\zeta$. Suppose that the optimal solution of the dual problem is ${\hat \zeta ^ * }$ and ${\Xi ^ * } \buildrel \Delta \over = \left\{ {\hat \alpha _{a\emph{i}}^ * ,\hat \alpha _{b\emph{i}}^ * } \right\}$. Then, we will discuss the solution of the dual problem in two cases.

For the first case, the optimal solution of the dual problem satisfies $\sum\limits_{U \in \left\{ {a,b} \right\}} {\sum\limits_{\emph{i} = 1}^K {\left( {{\hat \alpha _{U\emph{i}}^ *} - \hat \alpha _{U\emph{i}}^{ * 2}} \right)} } = 0$, notably, it is also a feasible solution to the original problem ${{\bf{P}}_{4.21}}$. Substituting ${\Xi ^ * }$ into ${{\bf{P}}_{4.21}}$, we obtain the following inequalities
\begin{align}
{d^ * } & \le \sum\limits_{\emph{i} = 1}^K {-\left( { R{{_s^{{\tau _\emph{i}}}}^\prime } + {\alpha _{b\emph{i}}} R{{_{s1}^{{t_\emph{i}}}}^\prime } + {\alpha _{a\emph{i}}}R{{_{s2}^{{t_\emph{i}}}}^\prime }} \right)}\notag \\
 &  = \mathcal{L}\left( {\hat \alpha _{ai}^ * ,\hat \alpha _{ai}^ * ,{{\hat \zeta }^ * }} \right) = {\varpi \left( {{{\hat \zeta }^*}} \right)} , \tag{D.2}\label{D2}
\end{align}
Comparing \eqref{D1} and \eqref{D2}, we have the following equality, which is given as
\begin{align}
&\mathop {\min }\limits_{\left( {{\alpha _{a\emph{i}}},{\alpha _{b\emph{i}}}} \right) \in \mathcal{D}} \mathop {\max }\limits_{\zeta  \ge 0} {\rm{ }}\;\;{\mathcal{L}}  \left( {{\alpha _{ai}},{\alpha _{bi}},\zeta } \right) \notag \\
 &= \mathop {\max }\limits_{\zeta  \ge 0}\mathop {\min }\limits_{\left( {{\alpha _{a\emph{i}}},{\alpha _{b\emph{i}}}} \right) \in \mathcal{D}} {\rm{ }}\;\;{\mathcal{L}}  \left( {{\alpha _{ai}},{\alpha _{bi}},\zeta } \right).\tag{D.3}\label{D4}
\end{align}
That's to say, when $\sum\limits_{U \in \left\{ {a,b} \right\}} {\sum\limits_{\emph{i} = 1}^K {\left( {{\hat \alpha _{U\emph{i}}^ *} - \hat \alpha _{U\emph{i}}^{ * 2}} \right)} } = 0$ holds, although ${\varpi \left( \zeta  \right)}$ is a monotonically increasing function with respect to $\zeta$, for $\forall \zeta  > {{\hat \zeta }^*}$, we have $\varpi \left( \zeta  \right) = {d^ * }$. That's is, the strong duality always holds for $\mathop {\min }\limits_{\left( {{\alpha _{a\emph{i}}},{\alpha _{b\emph{i}}}} \right) \in \mathcal{D}} \mathop {\max }\limits_{\zeta  \ge 0} {\rm{ }}\;\;{\mathcal{L}}  \left( {{\alpha _{ai}},{\alpha _{bi}},\zeta } \right)$ and $\mathop {\max }\limits_{\zeta  \ge 0}\mathop {\min }\limits_{\left( {{\alpha _{a\emph{i}}},{\alpha _{b\emph{i}}}} \right) \in \mathcal{D}} {\rm{ }}\;\;{\mathcal{L}}  \left( {{\alpha _{ai}},{\alpha _{bi}},\zeta } \right)$.

For the second case, i.e., $\sum\limits_{U \in \left\{ {a,b} \right\}} {\sum\limits_{\emph{i} = 1}^K {\left( {{\hat \alpha _{U\emph{i}}^ *} - \hat \alpha _{U\emph{i}}^{ * 2}} \right)} } > 0$. When $\zeta  > {{\hat \zeta }^*}$, $\mathop {\max }\limits_{\zeta  > 0} {\rm{ }}\varpi \left( \zeta  \right) \to \infty $ is unbounded due to the monotonically increasing nature of $\varpi \left( \zeta  \right)$, which contracts to the inequality in \eqref{D1} since the original problem in ${\bf{P}}_{4.21}$ must has a finite objective value. Based on the above, the $\sum\limits_{U \in \left\{ {a,b} \right\}} {\sum\limits_{\emph{i} = 1}^K {\left( {{\hat \alpha _{U\emph{i}}^ *} - \hat \alpha _{U\emph{i}}^{ * 2}} \right)} } = 0$ holds for the optimal solution and the proof is completed.

\end{appendices}

\ifCLASSOPTIONcaptionsoff
  \newpage
\fi
\bibliographystyle{IEEEtran}
\bibliography{ref}

\begin{thebibliography}{10}
\providecommand{\url}[1]{#1}
\csname url@samestyle\endcsname
\providecommand{\newblock}{\relax}
\providecommand{\bibinfo}[2]{#2}
\providecommand{\BIBentrySTDinterwordspacing}{\spaceskip=0pt\relax}
\providecommand{\BIBentryALTinterwordstretchfactor}{4}
\providecommand{\BIBentryALTinterwordspacing}{\spaceskip=\fontdimen2\font plus
\BIBentryALTinterwordstretchfactor\fontdimen3\font minus
  \fontdimen4\font\relax}
\providecommand{\BIBforeignlanguage}[2]{{%
\expandafter\ifx\csname l@#1\endcsname\relax
\typeout{** WARNING: IEEEtran.bst: No hyphenation pattern has been}%
\typeout{** loaded for the language `#1'. Using the pattern for}%
\typeout{** the default language instead.}%
\else
\language=\csname l@#1\endcsname
\fi
#2}}
\providecommand{\BIBdecl}{\relax}
\BIBdecl

\bibitem{8879484}
L.~Chettri and R.~Bera, ``A comprehensive survey on {I}nternet of {T}hings
  {(IoT)} toward 5{G} wireless systems,'' \emph{IEEE Internet Things J.},
  vol.~7, no.~1, pp. 16--32, 2020.

\bibitem{9475506}
N.~Khalid, R.~Mirzavand, H.~Saghlatoon, M.~M. Honari, A.~K. Iyer, and
  P.~Mousavi, ``A batteryless {RFID} sensor architecture with distance
  ambiguity resolution for smart home {IoT} applications,'' \emph{IEEE Internet
  Things J.}, vol.~9, no.~4, pp. 2960--2972, 2022.

\bibitem{10130082}
T.~Jiang, Y.~Zhang, W.~Ma, M.~Peng, Y.~Peng, M.~Feng, and G.~Liu, ``Backscatter
  communication meets practical battery-free {I}nternet of {T}hings: A survey
  and outlook,'' \emph{IEEE Commun. Surveys Tuts.}, vol.~25, no.~3, pp.
  2021--2051, 2023.

\bibitem{10437703}
H.~Guo, Y.~Ye, H.~Sun, and L.~Shi, ``Resource allocation for mutualistic
  symbiotic radio with hybrid active-passive communications,'' in \emph{Proc.
  IEEE Global Commun. Conf. {(GLOBECOM)}}, 2023, pp. 4418--4423.

\bibitem{8907447}
R.~Long, Y.-C. Liang, H.~Guo, G.~Yang, and R.~Zhang, ``Symbiotic radio: A new
  communication paradigm for passive {I}nternet of {T}hings,'' \emph{IEEE
  Internet Things J.}, vol.~7, no.~2, pp. 1350--1363, 2020.

\bibitem{9193946}
Y.-C. Liang, Q.~Zhang, E.~G. Larsson, and G.~Y. Li, ``Symbiotic radio:
  Cognitive backscattering communications for future wireless networks,''
  \emph{IEEE Trans. Cognit. Commun. Netw.}, vol.~6, no.~4, pp. 1242--1255,
  2020.

\bibitem{9749195}
Y.-C. Liang, R.~Long, Q.~Zhang, and D.~Niyato, ``Symbiotic communications:
  Where marconi meets darwin,'' \emph{IEEE Wireless Commun.}, vol.~29, no.~1,
  pp. 144--150, 2022.

\bibitem{11020590}
R.~Xu, Y.~Ye, H.~Sun, L.~Shi, and G.~Lu, ``Revolutionizing symbiotic radio:
  Exploiting trade-offs in hybrid active-passive communications,'' \emph{IEEE
  Commun. Mag.}, vol.~63, no.~9, pp. 156--163, 2025.

\bibitem{8692391}
H.~Guo, Y.-C. Liang, R.~Long, and Q.~Zhang, ``Cooperative ambient backscatter
  system: A symbiotic radio paradigm for passive {IoT},'' \emph{IEEE Wireless
  Commun. Lett.}, vol.~8, no.~4, pp. 1191--1194, 2019.

\bibitem{9866050}
Y.~Ye, L.~Shi, X.~Chu, G.~Lu, and S.~Sun, ``Mutualistic cooperative ambient
  backscatter communications under hardware impairments,'' \emph{IEEE Trans.
  Commun.}, vol.~70, no.~11, pp. 7656--7668, 2022.

\bibitem{8665892}
H.~Guo, Y.-C. Liang, R.~Long, S.~Xiao, and Q.~Zhang, ``Resource allocation for
  symbiotic radio system with fading channels,'' \emph{IEEE Access}, vol.~7,
  pp. 34\,333--34\,347, 2019.

\bibitem{9036977}
Z.~Chu, W.~Hao, P.~Xiao, M.~Khalily, and R.~Tafazolli, ``Resource allocations
  for symbiotic radio with finite blocklength backscatter link,'' \emph{IEEE
  Internet Things J.}, vol.~7, no.~9, pp. 8192--8207, 2020.

\bibitem{9686018}
J.~Xu, Z.~Dai, and Y.~Zeng, ``Enabling full mutualism for symbiotic radio with
  massive backscatter devices,'' in \emph{Proc. IEEE Global Commun. Conf.
  {(GLOBECOM)}}, 2021, pp. 1--6.

\bibitem{9461158}
H.~Yang, Y.~Ye, K.~Liang, and X.~Chu, ``Energy efficiency maximization for
  symbiotic radio networks with multiple backscatter devices,'' \emph{IEEE Open
  J. Commun. Soc.}, vol.~2, pp. 1431--1444, 2021.

\bibitem{9600844}
Y.~Guo, G.~Wang, R.~Xu, R.~He, X.~Wei, and C.~Tellambura, ``Capacity analysis
  for wireless symbiotic communication systems with {BPSK} tags under
  sensitivity constraint,'' \emph{IEEE Commun. Lett.}, vol.~26, no.~1, pp.
  44--48, 2022.

\bibitem{10214560}
W.~Zhao, J.~Zhu, X.~She, G.~Wang, and C.~Tellambura, ``Ergodic capacity
  analysis for cooperative ambient communication system under sensitivity
  constraint,'' \emph{IEEE Commun. Lett.}, vol.~27, no.~10, pp. 2822--2826,
  2023.

\bibitem{10130794}
J.~Wang, X.~Ding, Q.~Zhang, and Y.-C. Liang, ``Multiple access design for
  symbiotic radios: Facilitating massive {IoT} connections with cellular
  networks,'' \emph{IEEE Trans. Wireless Commun.}, vol.~23, no.~1, pp.
  201--216, 2024.

\bibitem{10107747}
M.~Ataeeshojai, R.~C. Elliott, W.~A. Krzymień, C.~Tellambura, and
  I.~Maljević, ``Symbiotic backscatter communication underlying a cell-free
  massive {MIMO} system,'' \emph{IEEE Internet Things J.}, vol.~10, no.~19, pp.
  16\,758--16\,777, 2023.

\bibitem{wang2017fm}
A.~Wang, V.~Iyer, V.~Talla, J.~R. Smith, and S.~Gollakota, ``{FM} backscatter:
  Enabling connected cities and smart fabrics,'' in \emph{Pro. 14th {USENIX}
  {S}ymp. {N}etw. {S}yst. {D}esign {I}mplementation {(NSDI)}}, 2017, pp.
  243--258.

\bibitem{zhang2016hitchhike}
P.~Zhang, D.~Bharadia, K.~Joshi, and S.~Katti, ``Hitchhike: Practical
  backscatter using commodity {W}i{F}i,'' in \emph{Pro. 14th {ACM} {C}onf.
  {E}mbedded {N}etwork {S}ensor {S}yst. {CD-ROM}}, 2016, pp. 259--271.

\bibitem{10463656}
M.~M. Butt, N.~R. Mangalvedhe, N.~K. Pratas, J.~Harrebek, J.~Kimionis,
  M.~Tayyab, O.-E. Barbu, R.~Ratasuk, and B.~Vejlgaard, ``{A}mbient {IoT}: A
  missing link in 3{GPP} {I}o{T} devices landscape,'' \emph{IEEE Internet
  Things Mag.}, vol.~7, no.~2, pp. 85--92, 2024.

\bibitem{8340034}
N.~Van~Huynh, D.~T. Hoang, D.~Niyato, P.~Wang, and D.~I. Kim, ``Optimal time
  scheduling for wireless-powered backscatter communication networks,''
  \emph{IEEE Wireless Commun. Lett.}, vol.~7, no.~5, pp. 820--823, 2018.

\bibitem{8957679}
B.~Lyu and D.~T. Hoang, ``Optimal time scheduling in relay assisted batteryless
  {I}o{T} networks,'' \emph{IEEE Wireless Commun. Lett.}, vol.~9, no.~5, pp.
  706--710, 2020.

\bibitem{9253590}
Y.~Zhuang, X.~Li, H.~Ji, H.~Zhang, and V.~C.~M. Leung, ``Optimal resource
  allocation for {RF}-powered underlay cognitive radio networks with ambient
  backscatter communication,'' \emph{IEEE Trans. Veh. Technol}, vol.~69,
  no.~12, pp. 15\,216--15\,228, 2020.

\bibitem{8327597}
B.~Lyu, H.~Guo, Z.~Yang, and G.~Gui, ``Throughput maximization for hybrid
  backscatter assisted cognitive wireless powered radio networks,'' \emph{IEEE
  Internet Things J.}, vol.~5, no.~3, pp. 2015--2024, 2018.

\bibitem{9154299}
R.~Long, Y.-C. Liang, Y.~Pei, and E.~G. Larsson, ``Active-load assisted
  symbiotic radio system in cognitive radio network,'' in \emph{IEEE 21st Int.
  Workshop on Signal Process. Adv. Wireless Commun.}, 2020, pp. 1--5.

\bibitem{9751388}
Q.~Zhang, Y.-C. Liang, H.-C. Yang, and H.~V. Poor, ``Mutualistic mechanism in
  symbiotic radios: When can the primary and secondary transmissions be
  mutually beneficial?'' \emph{IEEE Trans. Wireless Commun.}, vol.~21, no.~10,
  pp. 8036--8050, 2022.

\bibitem{9161012}
Y.~Ye, L.~Shi, X.~Chu, and G.~Lu, ``Throughput fairness guarantee in wireless
  powered backscatter communications with {HTT},'' \emph{IEEE Wireless Commun.
  Lett.}, vol.~10, no.~3, pp. 449--453, 2021.

\bibitem{razaviyayn2014successive}
M.~Razaviyayn, ``Successive convex approximation: Analysis and applications,''
  Ph.D. dissertation, University of Minnesota, 2014.

\bibitem{boyd2004convex}
S.~P. Boyd and L.~Vandenberghe, \emph{Convex optimization}.\hskip 1em plus
  0.5em minus 0.4em\relax Cambridge university press, 2004.

\bibitem{8571319}
D.~Xu and Q.~Li, ``Resource allocation for secure communications in cooperative
  cognitive wireless powered communication networks,'' \emph{IEEE Syst. J.},
  vol.~13, no.~3, pp. 2431--2442, 2019.

\bibitem{7981380}
S.~H. Kim and D.~I. Kim, ``Hybrid backscatter communication for
  wireless-powered heterogeneous networks,'' \emph{IEEE Trans. Wireless
  Commun.}, vol.~16, no.~10, pp. 6557--6570, 2017.

\bibitem{9534878}
Z.~Liu, S.~Zhao, Y.~Yang, K.~Ma, and X.~Guan, ``Toward hybrid backscatter-aided
  wireless-powered {I}nternet of {T}hings networks: Cooperation and coexistence
  scenarios,'' \emph{IEEE Internet Things J.}, vol.~9, no.~8, pp. 6264--6276,
  2022.

\bibitem{9151196}
Z.~Ding, R.~Schober, and H.~V. Poor, ``Unveiling the importance of {SIC} in
  {NOMA} systems—part 1: State of the art and recent findings,'' \emph{IEEE
  Commun. Lett.}, vol.~24, no.~11, pp. 2373--2377, 2020.

\bibitem{7959539}
Y.~Gao, B.~Xia, K.~Xiao, Z.~Chen, X.~Li, and S.~Zhang, ``Theoretical analysis
  of the dynamic decode ordering {SIC} receiver for uplink {NOMA} systems,''
  \emph{IEEE Commun. Lett.}, vol.~21, no.~10, pp. 2246--2249, 2017.

\bibitem{horst1999dc}
R.~Horst and N.~V. Thoai, ``{D}{C} programming: overview,'' \emph{J. Optim.
  Theory and Appl.}, vol. 103, pp. 1--43, 1999.

\bibitem{6698281}
U.~Rashid, H.~D. Tuan, H.~H. Kha, and H.~H. Nguyen, ``Joint optimization of
  source precoding and relay beamforming in wireless {MIMO} relay networks,''
  \emph{IEEE Trans. Commun.}, vol.~62, no.~2, pp. 488--499, 2014.

\bibitem{6816086}
E.~Che, H.~D. Tuan, and H.~H. Nguyen, ``Joint optimization of cooperative
  beamforming and relay assignment in multi-user wireless relay networks,''
  \emph{IEEE Trans. Wireless Commun.}, vol.~13, no.~10, pp. 5481--5495, 2014.

\bibitem{gondzio1994computational}
J.~Gondzio and T.~Terlaky, \emph{A computational view of interior-point methods
  for linear programming}.\hskip 1em plus 0.5em minus 0.4em\relax Citeseer,
  1994.

\bibitem{9222571}
G.~Yang, R.~Dai, and Y.-C. Liang, ``Energy-efficient uav backscatter
  communication with joint trajectory design and resource optimization,''
  \emph{IEEE Trans. Wireless Commun.}, vol.~20, no.~2, pp. 926--941, 2021.

\end{thebibliography}
\end{document}